\documentclass[12pt]{article}
\usepackage{amsmath,amssymb}
\usepackage{bm}
\usepackage{graphicx,color}
\usepackage{wrapfig}
\begin{document}
\title{
\begin{flushright}
\ \\*[-80pt] 
\begin{minipage}{0.2\linewidth}
\normalsize
%arXiv:YYMM.NNNN \\
\end{minipage}
\end{flushright}
{\Large \bf Predicting CP violation in  Deviation from Tri-bimaximal mixing of Neutrinos
\\*[20pt]}}

\author{
\centerline{Yusuke~Shimizu$^{1,}$\footnote{E-mail address: yusuke.shimizu@mpi-hd.mpg.de} \   ,
Morimitsu~Tanimoto$^{2,}$\footnote{E-mail address: tanimoto@muse.sc.niigata-u.ac.jp} \ \ 
and \ \ Kei Yamamoto$^{2,}$\footnote{E-mail address: yamamoto@muse.sc.niigata-u.ac.jp}}
\\*[20pt]
\centerline{
\begin{minipage}{\linewidth}
\begin{center}
$^1${\it \normalsize
\it \normalsize \it \normalsize Max-Planck-Institute f\"ur Kernphysik,
%Saupfercheckweg 1,
 D-69117 Heidelberg, Germany}
\\*[4pt]
$^2${\it \normalsize
Department of Physics, Niigata University,~Niigata 950-2181, Japan }
\end{center}
\end{minipage}}
\\*[70pt]}

\date{
\centerline{\small \bf Abstract}
\begin{minipage}{0.9\linewidth}
\vskip  1 cm
\small
We study the CP violation  in the deviation from the tri-bimaximal mixing (TBM) of neutrinos.
We examine non-trivial relations among  the mixing angles and the CP violating Dirac phase 
in the typical  four cases of the deviation from the TBM.
The first two cases are derived by the  additional rotation of the 2-3 or 1-3 generations of neutrinos in the TBM basis.
The other two cases are given 
by the additional rotation of the 1-3 or 1-2 generations of   charged leptons with the TBM neutrinos.
These four cases predict different relations  among three mixing angles and the CP violating Dirac phase.
The  rotation of the 2-3 generations of neutrinos in the TBM basis
  predicts $\sin ^2\theta _{12}<1/3$, and  the CP violating Dirac phase to be  $\pm (0.09\pi\sim 0.76\pi)$ for NH
 ($\pm (0.15\pi\sim 0.73\pi) \  \text{for IH}$) 
depending on $\sin ^2\theta _{23}$. 
The  rotation of the 1-3 generations of neutrinos in the TBM basis
 gives $\sin ^2\theta _{12}>1/3$.   The CP violating Dirac phase is not constrained by the input of the  present experimental data.
For the case of  the 1-3 and 1-2 rotations of charged leptons  in the TBM basis,
the CP violating Dirac phase is predicted in  $\pm(0.35\pi\sim 0.60\pi)$
depending on $\sin ^2\theta _{12}$ for both NH and IH cases.
We also discuss  the  specific case that   $\theta_{13}$ is related with the Cabibbo angle $\lambda$ such as $\sin\theta_{13}=\lambda/\sqrt{2}$,
in which the maximal CP violation is preferred.
The CP violating Dirac phase can distinguish the lepton flavor mixing patterns  at T2K and NO$\nu$A experiments in the future.
\end{minipage}
}

\begin{titlepage}
\maketitle
\thispagestyle{empty}
\end{titlepage}

\section{Introduction}

The neutrino oscillation experiments
 have  been revealed the neutrino masses and the two large neutrino  mixing angles. 
Reactor neutrino experiments
 have also observed a non-zero $\theta _{13}$, which is the last mixing angle of 
the lepton sector~\cite{An:2012eh,Ahn:2012nd,Adamson:2013whj,Abe:2014lus}.  
The T2K experiment has confirmed the neutrino oscillation in the $\nu_\mu \to\nu_e$ appearance events 
\cite{Abe:2013hdq,Abe:2014ugx},
which provide us the new information of the CP violation of the lepton sector by combining  the data of reactor experiments.
Thus,  recent neutrino oscillation experiments go into a new phase 
of precise determinations of the lepton mixing matrix and neutrino mass squared
 differences~\cite{Tortola:2012te,Fogli:2012ua,GonzalezGarcia:2012sz}. 
Therefore, the detailed study of the neutrino mixing including the CP violating phase
gives us clues to reach  the  flavor theory.

Before the reactor experiments reported 
the non-zero value of $\theta _{13}$, there was  a 
paradigm of "tri-bimaximal mixing" (TBM)~\cite{Harrison:2002er,Harrison:2002kp}, 
which is a simple mixing pattern for leptons and can be easily derived from flavor symmetries. 
At early stage some authors  succeeded to realize  the TBM in the $A_4$ models \cite{Ma:2001dn,Babu:2002dz,Altarelli:2005yp,Altarelli:2005yx,
deMedeirosVarzielas:2005qg}.
Then, the non-Abelian discrete groups are center of attention at the  flavor symmetry~
\cite{Altarelli:2010gt,Ishimori:2010au,Ishimori:2012zz}.
The other groups were also considered to give the TBM \cite{Mohapatra:2006pu}-\cite{Ishimori:2008fi}.
On the other hand, the deviation from the TBM was also discussed \cite{Xing:2002sw,Xing:2006ms}
 and the magnitudes of  $\theta _{13}$ was estimated based on models \cite{Adhikary:2006jx}-\cite{Antusch:2011qg}.
The observation of the non-vanishing $\theta_{13}$ forces to study the  deviation from 
the TBM~\cite{Ahn:2012tv}-\cite{King:2014nza} precisely or study other flavor paradigms,
e.g. tri-bimaximal-Cabibbo mixing~\cite{King:2012vj,Shimizu:2012ry}.

It is the important problem for the flavor physics
 whether the TBM basis has a physical background such as the flavor symmetry or not.  In order to answer this question,
 one should examine the flavor mixing angles and the CP violating Dirac phase strictly
 \cite{Marzocca:2013cr,Xing:2014zka,Petcov:2014uoa,Branco:2014zza,Kang:2014mka}.
In this work, we discuss predictions in the typical  four cases of the deviation from the TBM,
and propose to test the CP violating Dirac phase at the future experiments, T2K and NO$\nu$A.
The first two cases are given by the  additional rotation of 2-3 or 1-3 generations of neutrinos in the TBM basis,
%%%%%%%%%%%%%%%%%%%%%%%%%%%%%%%%%%%
% \footnote{A rotation of 1-2 generations still leads to $\theta _{13}=0$.}
%%%%%%%%%%%%%%%%%%%%%%%%%%%%%%%%%%%%
 which is called as the tri-maximal mixing \cite{Grimus:2008tt}.
Other two cases are given 
by the additional rotation of the 1-3 or 1-2 generations of charged leptons in the TBM basis, 
%%%%%%%%%%%%%%%%%%%%%%%%%
%\footnote{A rotation of 2-3 generations also gives $\theta _{13}=0$.}, 
%%%%%%%%%%%%%%%%%%%%%%%%%
which are not any more the tri-maximal mixing.
These four cases give  different relations  among three mixing angles and the CP violating Dirac phase.

%the magnitude of $\sin ^2\theta _{12}$. 
%The additional 1-3 rotation strictly leads to $\sin ^2\theta _{12}>1/3$. 
%On the other hand, in the case of 2-3 rotation, 
%the additional rotation can give rise to $\sin ^2\theta _{12}<1/3$, 
%which is in favor for experimental results. 
%Therefore, we will focus on the additional 2-3 rotation to the TBM. 

First, we discuss the  rotation of the $2$-$3$ generations of neutrinos in the TBM basis, which
  predicts $\sin ^2\theta _{12}<1/3$.  
It is also found that the CP violating Dirac phase $\delta _{CP}$ is non-zero and  depends on 
the magnitude of $\sin^2\theta _{23}$.
The case of  the  rotation of the $1$-$3$ generations of neutrinos in the TBM basis
 gives $\sin ^2\theta _{12}>1/3$.  The CP violating Dirac phase is not constrained by the input of the present experimental data.
For the case of the  rotation of the 1-3 and 2-3 generations of charged leptons   in the TBM basis,
   $\sin ^2\theta _{23}$ is very close to $1/2$ with larger than  $1/2$ (1-3 rotation) or smaller than  $1/2$ (1-2 rotation).
For both cases, the CP violating Dirac phase is predicted to be in the narrow range of
 $\pm (0.35\pi\sim 0.60\pi)$,
which may be  preferred by the recent T2K experiment data 
\cite{Abe:2013hdq,Abe:2014ugx}.
Thus, the CP violating Dirac phase can distinguish the lepton flavor mixing patterns deviating from the TBM.

In addition,  we discuss the specific case where the $\theta_{13}$ is related with the Cabibbo angle.  In the framework of Grand Unified Theories (GUT), where the Yukawa matrices for the charged leptons and the down-type quarks have the same origin,   the $1$-$2$ generation mixing 
angle of the charged leptons is  same as  the Cabibbo angle $\lambda$ 
 \cite{King:2012vj,Antusch:2013kna}. Therefore, one obtains $\sin\theta_{13}=\lambda/\sqrt{2}$,
 which is  in excellent  agreement with the observed value.
 Then, we  predict the magnitude of the  CP violating phase.

The paper is organized as follows. We discuss the non-trivial relations among mixing angles and the CP violating Dirac phase
in the  four cases of the deviation from the TBM in the section~2. 
In the section~3, we present numerical predictions.
 The section 4 is devoted to the summary.

%%%%%%%%%%%%%%%%%%%%%%
\section{Deviation from tri-bimaximal mixing}
Let us discuss the structure of the lepton mixing matrix, so called the 
Pontecorvo-Maki-Nakagawa-Sakata (PMNS) matrix $U_{\text{PMNS}}$~\cite{Maki:1962mu,Pontecorvo:1967fh}.
Three mixing angles $\theta _{ij}$ $(i,j=1,2,3;~i<j)$ and one CP violating Dirac phase $\delta _{CP}$
are parameterized as  the PDG form~\cite{Beringer:1900zz}:
\begin{align}
U_\text{PMNS} \equiv 
%\begin{pmatrix}
%1 & 0 & 0 \\
%0 & c_{23} & s_{23} \\
%0 & -s_{23} & c_{23}
%\end{pmatrix} 
%\begin{pmatrix}
%c_{13} & 0 & s_{13}e^{-i\delta _{CP}} \\
%0 & 1 & 0 \\
%-s_{13}e^{i\delta _{CP}} & 0 & c_{13}
%\end{pmatrix}
%\begin{pmatrix}
%c_{12} & s_{12} & 0 \\
%-s_{12} & c_{12} & 0 \\
%0 & 0 & 1
%\end{pmatrix} \nonumber \\
%= 
\begin{pmatrix}
c_{12} c_{13} & s_{12} c_{13} & s_{13}e^{-i\delta _{CP}} \\
-s_{12} c_{23} - c_{12} s_{23} s_{13}e^{i\delta _{CP}} & 
c_{12} c_{23} - s_{12} s_{23} s_{13}e^{i\delta _{CP}} & s_{23} c_{13} \\
s_{12} s_{23} - c_{12} c_{23} s_{13}e^{i\delta _{CP}} & 
-c_{12} s_{23} - s_{12} c_{23} s_{13}e^{i\delta _{CP}} & c_{23} c_{13}
\end{pmatrix},
\label{UPMNS}
\end{align}
where $c_{ij}$ and  $s_{ij}$ denote $\cos \theta _{ij}$ and $\sin \theta _{ij}$, respectively. 
If neutrinos are Majorana particles, Majorana phases are included 
in the left-handed Majorana neutrino masses. As well known the Jarlskog invariant~\cite{Jarlskog:1985ht}, 
which is the  parameter
describing the size of the CP violation, is written as 
\begin{equation}
J_{CP}=\text{Im}\left [U_{e1}U_{\mu 2}U_{e2}^\ast U_{\mu 1}^\ast \right ],
\end{equation}
where $U_{ij}$s are the PMNS matrix elements. 
Then, the $J_{CP}$ is written in terms of the  lepton mixing angles and the CP violating Dirac phase as
\begin{equation}
J_{CP}=s_{23}c_{23}s_{12}c_{12}s_{13}c_{13}^2\sin \delta _{CP}~.
\end{equation}

For the lepton mixing matrix, Harrison-Perkins-Scott proposed a simple form of the mixing pattern, 
so-called the tri-bimaximal mixing~\cite{Harrison:2002er,Harrison:2002kp} as follows:
\begin{equation}
V_{\text{TBM}}=
\begin{pmatrix}
\frac{2}{\sqrt{6}} & \frac{1}{\sqrt{3}} & 0 \\
-\frac{1}{\sqrt{6}} & \frac{1}{\sqrt{3}} & -\frac{1}{\sqrt{2}} \\
-\frac{1}{\sqrt{6}} & \frac{1}{\sqrt{3}} & \frac{1}{\sqrt{2}}
\end{pmatrix}.
\label{UTBM}
\end{equation}
Before the reactor experiments reported non-zero $\theta _{13}$ \cite{An:2012eh,Ahn:2012nd,Adamson:2013whj,Abe:2014lus}, 
the TBM  was good scheme for the lepton sector. 
In our work, we assume the TBM to be a good starting point of the  lepton mixing.
Then, we discuss the deviation from the TBM, 
 which originates in the neutrino sector or in the charged lepton sector.
These different origins give us the different flavor mixing patterns.

%%%%%%%%%%%%%%%%%%%%%%%%%%%%%%%%%%%%%%%%%%%%%%%%%%%%%%%%%%%%%%%%%%%%%%%%%%%%%%%%%%%%%%%%%%
\subsection{Deviation in the neutrino sector}
First of all, we consider an additional rotation of $2$-$3$ generations of neutrinos
in the TBM basis ~\cite{Rodejohann:2012cf,Shimizu:2012ry},
which is called as Case I.
In order to obtain the PMNS mixing matrix, we  multiply the TBM matrix $V_{\text{TBM}}$ by the 2-3 rotation matrix in the right-hand side  as follows:
\begin{equation}
U_{\text{PMNS}}=V_{\text{TBM}}
\begin{pmatrix}
1 & 0 & 0 \\
0 & \cos \phi & e^{-i\sigma }\sin \phi \\
0 & -e^{i\sigma }\sin \phi & \cos \phi 
\end{pmatrix},
\label{RO1}
\end{equation}
where $\phi$ and $\sigma$ are arbitrary parameters, which are determined by the experimental data.
Then, the relevant mixing matrix elements are given as
\begin{equation}
|U_{e2}|=\left |\frac{\cos \phi }{\sqrt{3}}\right |,\quad 
|U_{e3}|=\left |\frac{e^{-i\sigma }\sin \phi }{\sqrt{3}}\right |,\quad 
|U_{\mu 3}|=\left |-\frac{\cos \phi }{\sqrt{2}}+\frac{e^{-i\sigma }\sin \phi }{\sqrt{3}}\right |,
\end{equation}
which are converted  to
\begin{equation}
\sin ^2\theta _{12}=1-\frac{2}{3-\sin ^2\phi },\quad \sin ^2\theta _{13}=\frac{1}{3}\sin ^2\phi ,\quad 
\sin ^2\theta _{23}=\frac{1}{2}\left (1-\frac{\sqrt{6}\sin 2\phi \cos \sigma }{3-\sin ^2\phi }\right ).
\label{RO1mix}
\end{equation}
The Jarlskog invariant $J_{CP}$ is given as
\begin{equation}
J_{CP}=s_{23}c_{23}s_{12}c_{12}s_{13}c_{13}^2\sin \delta _{CP}~=-\frac{1}{6\sqrt{6}}\sin 2\phi \sin \sigma \ ,
\label{JCP}
\end{equation}
and then, the CP phase $\delta _{CP}$ is expressed in terms of two parameters  $\phi$ and $\sigma$ as follows:
\begin{equation}
\sin \delta _{CP}=-\frac{ (5+\cos 2\phi )\sin \sigma }
{\sqrt{(5+\cos 2\phi )^2-24\sin ^22\phi \cos ^2\sigma }}~.
\label{CP1}
\end{equation}

Since this  case does not change the first column of the TBM mixing matrix, 
this is called as TM1,  which is one of the  tri-maximal mixing patterns. 
Then, $\cos \delta _{CP}$ is easily obtained by putting $|U_{\mu 1}|^2=1/6$.
Now, we can write down the sum rules among the mixing angles and the phase $\delta_{CP}$
 from  Eqs. (\ref{RO1mix}) and (\ref{CP1}) \cite{Albright:2010ap}:
\begin{eqnarray}
\sin^2\theta_{12}=1-\frac{2}{3}\frac{1}{\cos^2\theta_{13}}\leq \frac{1}{3} \ , \qquad
\cos\delta_{CP}\tan 2\theta_{23}\simeq
  - \frac{1}{2\sqrt{2}\sin\theta_{13}}\left (1-\frac{7}{2}\sin^2\theta_{13}\right )  \ ,
\label{sum1}
\end{eqnarray}
 which are helpful to understand our numerical results in the section 3.

It is remarked that this mixing pattern is derived from the  neutrino mass matrix $m_{\nu LL}$
of   the $A_4$ and $S_4$ models \cite{Rodejohann:2012cf,Li:2013jya}:
\begin{equation}
m_{\nu LL}= a
\begin{pmatrix}
2&-1&-1\cr -1&2 &-1\cr -1&-1&2
\end{pmatrix} 
+b
\begin{pmatrix} 
1&0 &0\cr 0&0 &1\cr 0&1 &0
\end{pmatrix}+
c
\begin{pmatrix}
 0&1 &1\cr 1&1 &0\cr 1&0 &1
\end{pmatrix}
+d
\begin{pmatrix}
 0&1 &-1\cr 1&2 &0\cr -1&0 &-2
 \end{pmatrix} \ ,
 \end{equation}
where $a$, $b$, $c$ and $d$ are arbitrary complex parameters.

Next, we discuss  Case II, in which  an additional rotation of $1$-$3$ generations of neutrinos  is taken
in the TBM basis~\cite{Xing:2002sw,Ishimori:2010fs,Shimizu:2011xg}. 
The  PMNS matrix is given as 
\begin{equation}
U_{\text{PMNS}}=V_{\text{TBM}}
\begin{pmatrix}
\cos \phi & 0 & e^{-i\sigma }\sin \phi \\
0 & 1 & 0 \\
-e^{i\sigma }\sin \phi & 0 & \cos \phi 
\end{pmatrix}.
\label{RO2}
\end{equation}
The relevant mixing matrix elements are written as
\begin{equation}
|U_{e2}|=\frac{1}{\sqrt{3}},\quad 
|U_{e3}|=\left |\frac{2e^{-i\sigma }\sin \phi }{\sqrt{6}}\right |,\quad 
|U_{\mu 3}|=\left |-\frac{\cos \phi }{\sqrt{2}}-\frac{e^{-i\sigma }\sin \phi }{\sqrt{6}}\right |,
\end{equation}
which are converted  to
\begin{equation}
\sin ^2\theta _{12}=\frac{1}{3-2\sin ^2\phi },\quad \sin ^2\theta _{13}=\frac{2}{3}\sin ^2\phi ,\quad 
\sin ^2\theta _{23}=\frac{1}{2}\left (1+\frac{\sqrt{3}\sin 2\phi \cos \sigma }{3-2\sin ^2\phi }\right ).
\label{RO2mix}
\end{equation}
The Jarlskog invariant $J_{CP}$  is given as follows:
\begin{equation}
J_{CP}=-\frac{1}{6\sqrt{3}}\sin 2\phi \sin \sigma \ .
\end{equation}
Therefore, the CP phase $\delta _{CP}$ is given as
\begin{equation}
\sin \delta _{CP}=-\frac{ (2+\cos 2\phi )\sin \sigma }
{\sqrt{(2+\cos 2\phi )^2-3\sin^2 2\phi \cos^2 \sigma }}~.
\label{CP2}
\end{equation}

The rotation of the $1$-$3$ generations of neutrinos in the TBM basis  gives another
tri-maximal mixing pattern, in which 
the second column of the TBM mixing matrix is not changed,  so  called TM2. 
Then, $\cos \delta _{CP}$ is  obtained by putting $|U_{\mu 2}|^2=1/3$.
The sum rules among the mixing angles and the phase $\delta_{CP}$ are given by 
 from  Eqs. (\ref{RO2mix}) and (\ref{CP2}) \cite{Albright:2010ap}:
\begin{eqnarray}
\sin^2\theta_{12}=\frac{1}{3}\frac{1}{\cos^2\theta_{13}}\geq \frac{1}{3} \ , \qquad
\cos\delta_{CP}\tan 2\theta_{23}\simeq
  \frac{1}{\sqrt{2}\sin\theta_{13}} \left( 1-\frac{5}{4}\sin^2\theta_{13}\right ) \ .
\label{sum2}
\end{eqnarray}

 This mixing  pattern is obtained in the neutrino mass matrix $m_{\nu LL}$  of 
 the  $A_4$ and $S_4$ models \cite{Ishimori:2010fs,Shimizu:2011xg,Hagedorn:2012ut,King:2014nza,Ding:2013hpa}:
\begin{equation}
m_{\nu LL}= a
\begin{pmatrix}
2&-1&-1\cr -1&2 &-1\cr -1&-1&2
\end{pmatrix} 
+b
\begin{pmatrix} 
1&0 &0\cr 0&0 &1\cr 0&1 &0
\end{pmatrix}+
c
\begin{pmatrix}
 0&1 &1\cr 1&1 &0\cr 1&0 &1
\end{pmatrix}
+d
\begin{pmatrix}
 0&1 &-1\cr 1&-1 &0\cr -1&0 &1
 \end{pmatrix} \ ,
 \end{equation}
where $a$, $b$, $c$ and $d$ are arbitrary complex parameters.
 
Since the  rotation of $1$-$2$ generations leads to
$\theta _{13}=0$, we do not discuss this case.

%%%%%%%%%%%%%%%%%%%%%%%%%%%%%%%%%%%%%%%%%%%%%%%%%%%%%%%%%%%%%%%%%%%%%%%%%%%%%%%%%%%%%%%%%%%
\subsection{Deviation in the  charged lepton sector}
 The TBM  is realized in the specific flavor structure of the neutrino mass matrix  in the basis of the diagonal charged lepton mass matrix,
which is given in some flavor symmetries.
 If an additional  flavor symmetry breaking term causes 
the deviation from the diagonal charged lepton mass matrix, 
 the lepton mixing matrix is not anymore the tri-maximal mixing.
% By introducing an extra $\bf 3$ flavon, which couples only to the charged
% lepton mass matrix,
% in the $A_4$ model, we can easily get the   1-3 or  1-2 rotations  of the % diagonal charged lepton mass matrix
% in the good approximation.

Let us discuss Case III, in which the  neutrino mixing is the TBM one,
 on the other hand,
 the diagonal charged lepton mass matrix is rotated between $1$-$3$ generations.
In order to obtain the PMNS mixing matrix, we multiply the TBM matrix by the 1-3 rotation matrix 
in the left-hand side as follows:
\begin{equation}
U_{\text{PMNS}}=
\begin{pmatrix}
\cos \phi & 0 & -e^{-i\sigma }\sin \phi \\
0 & 1 & 0 \\
e^{i\sigma }\sin \phi & 0 & \cos \phi 
\end{pmatrix}V_{\text{TBM}} \ .
\label{RO3}
\end{equation}
Then, the relevant mixing matrix elements are given as
\begin{equation}
|U_{e2}|=\left |\frac{\cos \phi }{\sqrt{3}}-\frac{e^{-i\sigma }\sin \phi }{\sqrt{3}}\right |,\quad 
|U_{e3}|=\left |-\frac{e^{-i\sigma }\sin \phi }{\sqrt{2}}\right |,\quad 
|U_{\mu 3}|=\frac{1}{\sqrt{2}} \ ,
\label{RO3mix}
\end{equation}
which are converted to
\begin{equation}
\sin ^2\theta _{12}=\frac{2(1-\sin 2\phi \cos \sigma )}{3(2-\sin ^2\phi )},\quad 
\sin ^2\theta _{13}=\frac{1}{2}\sin ^2\phi ,\quad \sin ^2\theta _{23}=\frac{1}{2-\sin ^2\phi } \ .
\label{RO3mix2}
\end{equation}
The Jarlskog invariant $J_{CP}$ is given as
\begin{equation}
J_{CP}=\frac{1}{12}\sin 2\phi \sin \sigma \ .
\end{equation}
Therefore, the  CP phase $\delta _{CP}$ is written as 
\begin{equation}
\sin \delta _{CP}=\frac{(2-\sin ^2\phi )\sin \sigma }
{\sqrt{ (4-3\sin ^2\phi +2\sin 2\phi \cos \sigma )(1-\sin 2\phi \cos \sigma )}}~.
\label{CP3}
\end{equation}
On the other hand, $\cos \delta _{CP}$ is  obtained by putting $|U_{\mu 1}|^2=1/6$
as well as in  Case I.
The sum rules among the mixing angles and the phase $\delta_{CP}$ are given 
 from  Eqs. (\ref{RO3mix2}) and (\ref{CP3}) \cite{Albright:2010ap}:
\begin{eqnarray}
\sin^2\theta_{23}=\frac{1}{2}\frac{1}{\cos^2\theta_{13}}\geq \frac{1}{2} \ , \quad
\sin^2\theta_{12}\simeq
  \frac{1}{\sqrt{3}}
 -\frac{2\sqrt{2}}{3}\sin\theta_{13}\cos \delta_{CP}
 +\frac{1}{3} \sin^2\theta_{13} \cos 2\delta_{CP} \ .
\label{sum3}
\end{eqnarray}

Finally, we  discuss  Case IV,  in which the  neutrino mixing is the TBM one, on the other hand,
 the diagonal charged lepton mass matrix is rotated between $1$-$2$ generations.
The PMNS mixing matrix is given as 
\begin{equation}
U_{\text{PMNS}}=
\begin{pmatrix}
\cos \phi & -e^{-i\sigma }\sin \phi & 0 \\
e^{i\sigma }\sin \phi & \cos \phi & 0 \\
0 & 0 & 1
\end{pmatrix}V_{\text{TBM}}\ .
\label{RO4}
\end{equation}
The relevant mixing matrix elements are 
\begin{equation}
|U_{e2}|=\left |\frac{\cos \phi }{\sqrt{3}}-\frac{e^{-i\sigma }\sin \phi }{\sqrt{3}}\right |,\quad 
|U_{e3}|=\left |\frac{e^{-i\sigma }\sin \phi }{\sqrt{2}}\right |,\quad 
|U_{\mu 3}|=\left |-\frac{\cos \phi }{\sqrt{2}}\right |~,
\label{RO4mix}
\end{equation}
which are converted to
\begin{equation}
\sin ^2\theta _{12}=\frac{2(1-\sin 2\phi \cos \sigma )}{3(2-\sin ^2\phi )},\quad 
\sin ^2\theta _{13}=\frac{1}{2}\sin ^2\phi ,\quad \sin ^2\theta _{23}=1-\frac{1}{2-\sin ^2\phi } \ .
\label{RO4mix2}
\end{equation}
The Jarlskog invariant $J_{CP}$ is given as follows:
\begin{equation}
J_{CP}=-\frac{1}{12}\sin 2\phi \sin \sigma \ .
\end{equation}
The CP phase $\delta _{CP}$ is 
\begin{equation}
\sin \delta _{CP}=-\frac{ (2-\sin ^2\phi )\sin \sigma }
{\sqrt{(4-3\sin ^2\phi +2\sin 2\phi \cos \sigma )(1-\sin 2\phi \cos \sigma )}}~,
\label{CP4}
\end{equation}
and $\cos \delta _{CP}$ is given by putting $|U_{\tau 1}|^2=1/6$.

It is noticed that the mixing angles in Eq.(\ref{RO4mix2}) are obtained by
 replacing $\sin^2\theta _{23}$ with $\cos^2\theta _{23}$  in Eq.(\ref{RO3mix2}), and 
 $\sin \delta _{CP}$ is the same as the one of Case III except  $\pm$ sign.
%We also obtain  $\cos \delta_{CP}$  by
% replacing $\sin\theta _{23}$ with $\cos\theta _{23}$ in the one in Case III
%except  $\pm$ sign.
The sum rules among the mixing angles and the phase $\delta_{CP}$ are given by 
 from  Eqs.(\ref{RO4mix}) and (\ref{CP4}) \cite{Albright:2010ap}:
\begin{eqnarray}
\sin^2\theta_{23}=1-\frac{1}{2}\frac{1}{\cos^2\theta_{13}}\leq \frac{1}{2} , \quad
\sin^2\theta_{12}\simeq
  \frac{1}{\sqrt{3}}
 +\frac{2\sqrt{2}}{3}\sin\theta_{13}\cos \delta_{CP}
 +\frac{1}{3} \sin^2\theta_{13} \cos 2\delta_{CP} .
\label{sum4}
\end{eqnarray}

 We also discuss the specific case in Case IV, where  $\theta_{13}$ is related with the Cabibbo angle $\lambda$.  
In the framework of GUT, where the Yukawa matrices for the charged leptons and  the down-type quarks have the same origin,  the  $1$-$2$ generation mixing angle $\sin\phi$ is  the same
as  the Cabibbo angle $\lambda$
 \cite{King:2012vj,Antusch:2013kna}. Therefore, one obtains $\sin\theta_{13}=\lambda/\sqrt{2}\simeq 0.16$,
 which is  in excellent  agreement with the observed value.
 Then, the magnitude of the CP violating phase is predicted to be in the narrow range.

Since the  rotation of $2$-$3$ generations  gives
  $\theta _{13}=0$,  we do not discuss this case.
 By using above formulas, the numerical studies are discussed in the next section.

%%%%%%%%%%%%%%%%%%%%%%%%%%%%%%%%%%%%%%%%%%%%%%%%%%%%%%%%%%%%%%%%%%%%%%%%%%%%%%%
%%%%%%%%%%%%%%%%%%%    Numerical analysis   %%%%%%%%%%%%%%%%%%%%%%%%%%%%%%%%%%%
%%%%%%%%%%%%%%%%%%%%%%%%%%%%%%%%%%%%%%%%%%%%%%%%%%%%%%%%%%%%%%%%%%%%%%%%%%%%%%%
\section{Numerical analysis}
%%%%%%%%%%%%%%%%%%%%%%%%%%%%%%%%%%%%%%%%%%%%%
\begin{figure}[t!]
%\begin{minipage}[]{0.45\linewidth}
\vspace{4 mm}
\includegraphics[width=7.5cm]{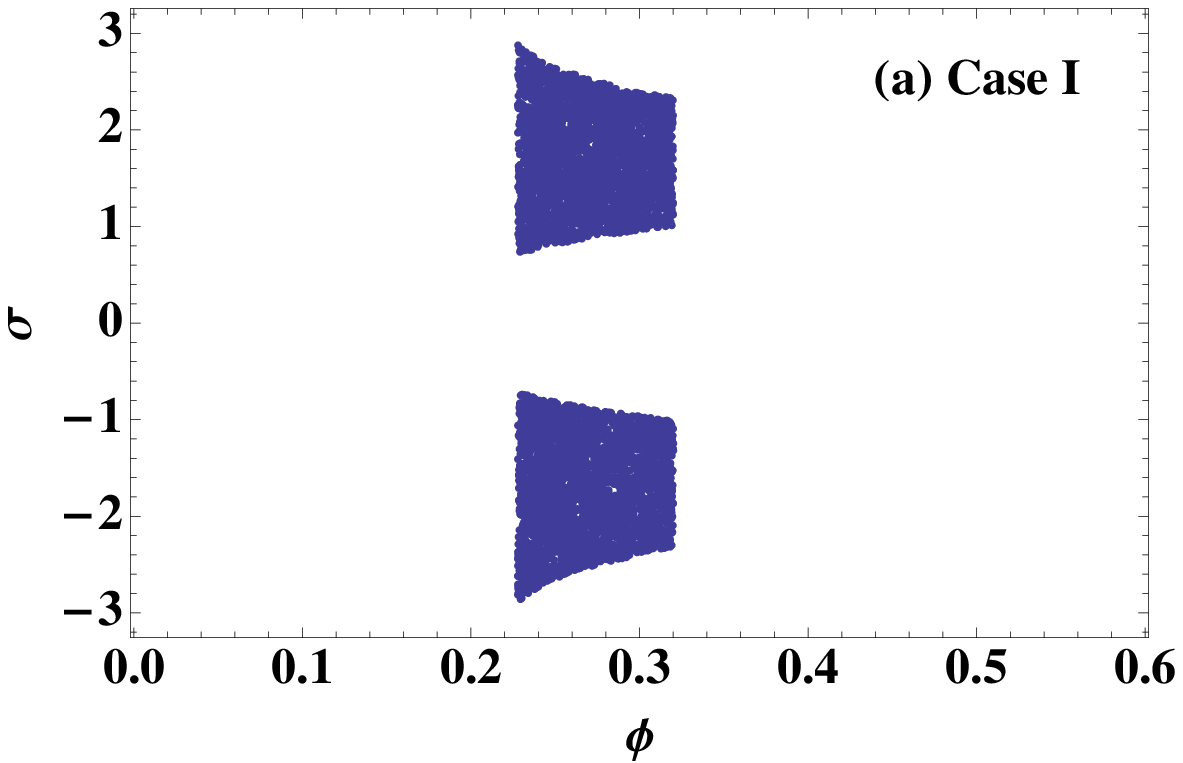}
%\end{minipage}
\hspace{5mm}
%\begin{minipage}[]{0.45\linewidth}
%\includegraphics[width=7.5cm]{theta13-delta-TBM-23.eps}
\includegraphics[width=7.5cm]{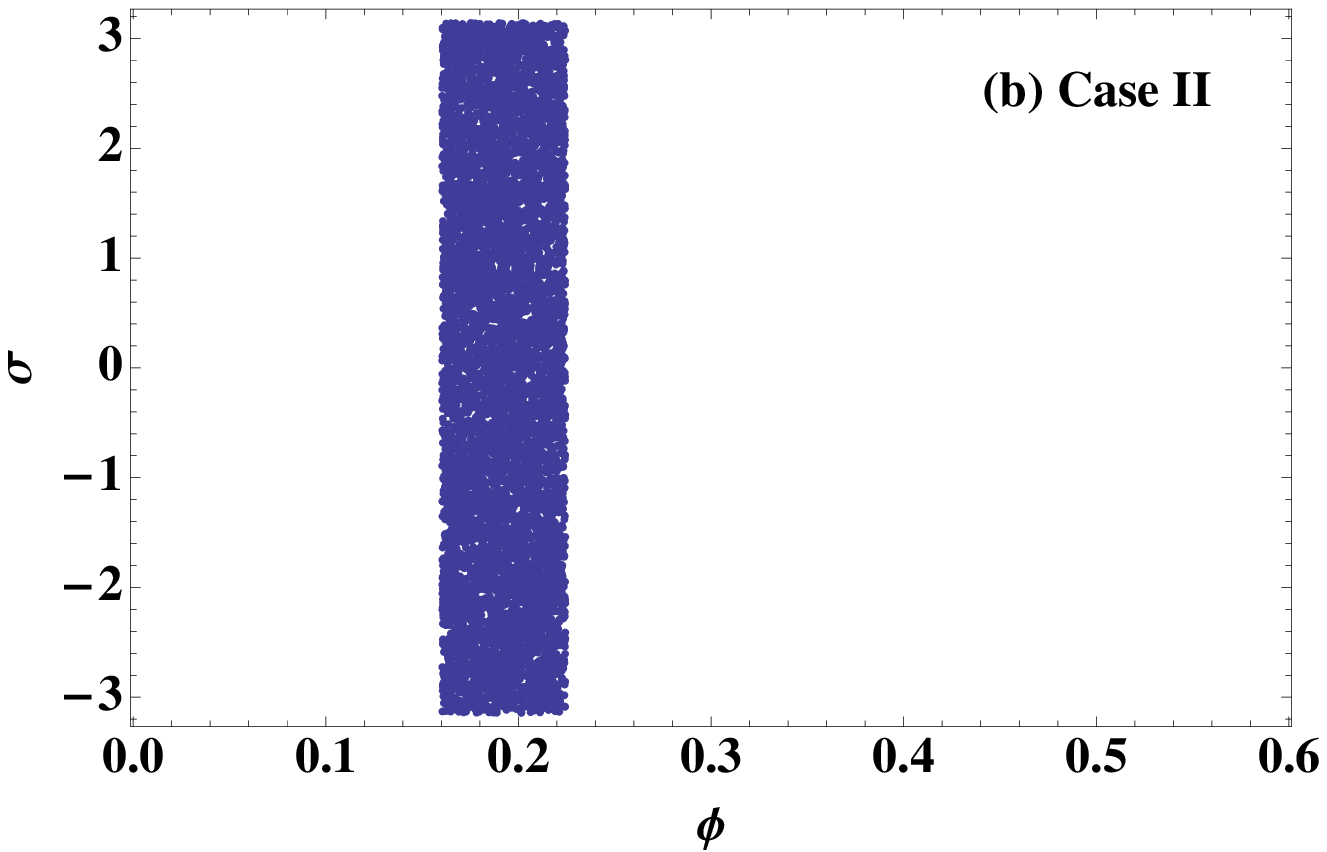}
\vskip 0.5 cm
\includegraphics[width=7.5cm]{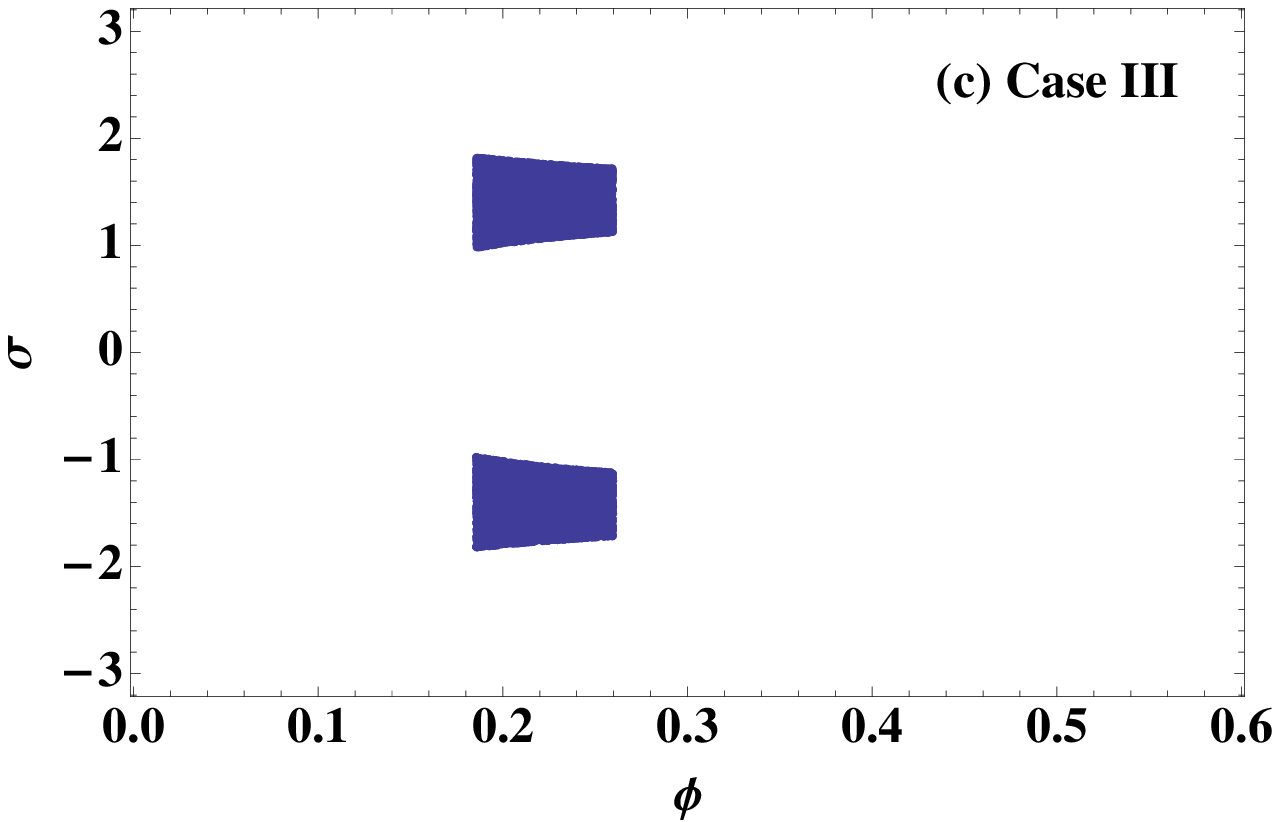}
\hspace{5 mm}
\includegraphics[width=7.5cm]{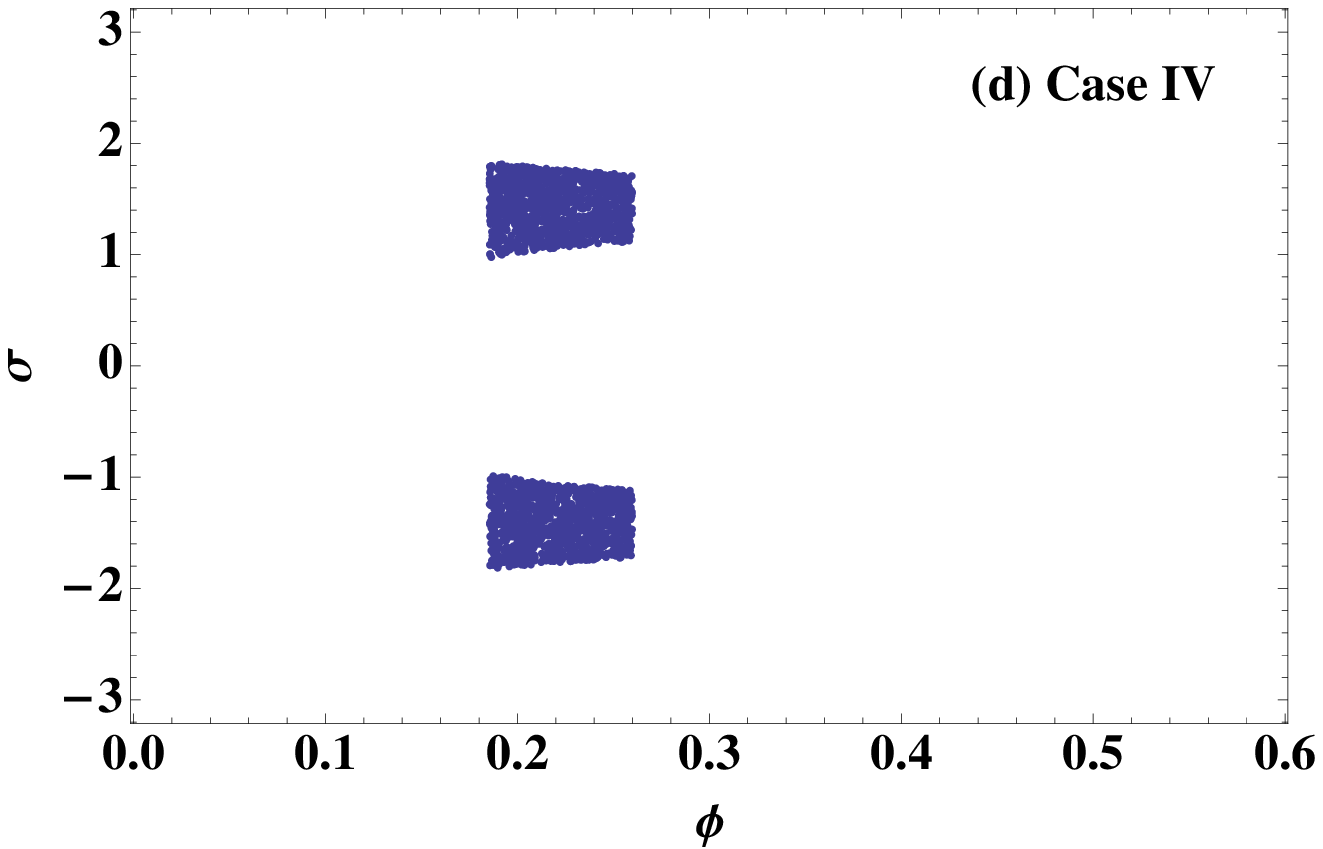}
\caption{Allowed regions  on $\phi$--$\sigma$ plane for NH
 in (a)  Case I,  (b)  Case II,  (c)  Case III and (d)  Case IV, where
 the experimental data  are input at $3\sigma$ level.
}
\label{fig1}
%\end{minipage}
\end{figure}
%%%%%%%%%%%%%%%%%%%%%%%%%%%%%%%%%%%%%%%%%%%%%
%%%%%%%%%%%%%%%%%%%%%%%%%%%%%%%%%%%%%%%%%%%%%
%%%%%%%%%%%%%%%%%%%%%%%%%%%%%%%%%%%%%%%%%%%%%
As input data in our calculations, we use the results of the global analysis of the neutrino oscillation experiments for  three mixing angles~\cite{Tortola:2012te}.
  For the case of the normal  hierarchy (NH) of neutrino masses,  we take
\begin{equation}
\sin ^2\theta _{12}=0.320_{-0.017}^{+0.016}~,\quad 
\sin ^2\theta _{13}=0.0246_{-0.0028}^{+0.0029}~,\quad 
\sin ^2\theta _{23}=0.613_{-0.040}^{+0.022}~(0.427_{-0.027}^{+0.034}),
%\footnote{This is a local minimum. The detail is shown in Ref.~\cite{Tortola:2012te}.}~,
\end{equation}
 at $1~\sigma $ level, where there are two solutions 
for $\sin ^2\theta _{23}$,  and 
\begin{equation}
0.27<\sin ^2\theta _{12}<0.37~,\quad 
0.017<\sin ^2\theta _{13}<0.033~,\quad 
0.36<\sin ^2\theta _{23}<0.68~,
\label{lepton-mixing-3sigma}
\end{equation}
at $3~\sigma $ level.
For  the case of  the inverted hierarchy (IH) of neutrino masses, we take 
\begin{equation}
\sin ^2\theta _{12}=0.320_{-0.017}^{+0.016}~,\quad 
\sin ^2\theta _{13}=0.0250_{-0.0027}^{+0.0026}~,\quad 
\sin ^2\theta _{23}=0.600_{-0.031}^{+0.026}~,
\end{equation}
at $1~\sigma $ level.  We have  at $3~\sigma $ level:
\begin{equation}
0.27<\sin ^2\theta _{12}<0.37~,\quad 
0.017<\sin ^2\theta _{13}<0.033~,\quad 
0.37<\sin ^2\theta _{23}<0.67~.
\label{IH3sigma}
\end{equation}

Our numerical procedure is as follows.
The  free parameters $\phi$ and $\sigma$ are scattered in the regions of
 $0 <\phi <\pi/2 $ and  $-\pi <\sigma <\pi $, respectively,
and then, we obtain  the regions of $\phi$  and $\sigma$ which are allowed by the experimental data of three mixing angles $\theta_{12}$,  $\theta_{23}$ and  $\theta_{13}$
 at $3\sigma$ level.
 Using these allowed sets $\phi$ and $\sigma$, we predict the CP violating phase $\delta_{CP}$
 by plugging back to, for example,  Eqs. (\ref{RO1mix}) and (\ref{CP1}) for Case I.
 We show the allowed regions of $\phi$  and $\sigma$ for the NH case  in Fig. 1.
 The allowed region in Case III and  Case IV are just same because
 the region of  $\phi$ is determined practically only by the experimental data
 of $\sin^2 \theta_{13}$, which is the same form as seen in  Eqs.(\ref{RO3mix2}) and (\ref{RO4mix2}),
 and $\sin \delta _{CP}$ is the same  one  except the $\pm$ sign
 as seen in Eqs. (\ref{CP3}) and (\ref{CP4}).
 
 The CP violating phase $\delta_{CP}$ is predicted by these allowed region in Fig.1.
 By using these allowed sets ($\phi, \sigma$), we show 
  relations among the mixing angles and the CP violating phase numerically.

%%%%%%%%%%%%%%%%%%%%%%%%%%%%%%%%%%%%%%%%%%%%%
\begin{figure}[b!]
\begin{minipage}[]{0.45\linewidth}
\vspace{4 mm}
\includegraphics[width=7.5cm]{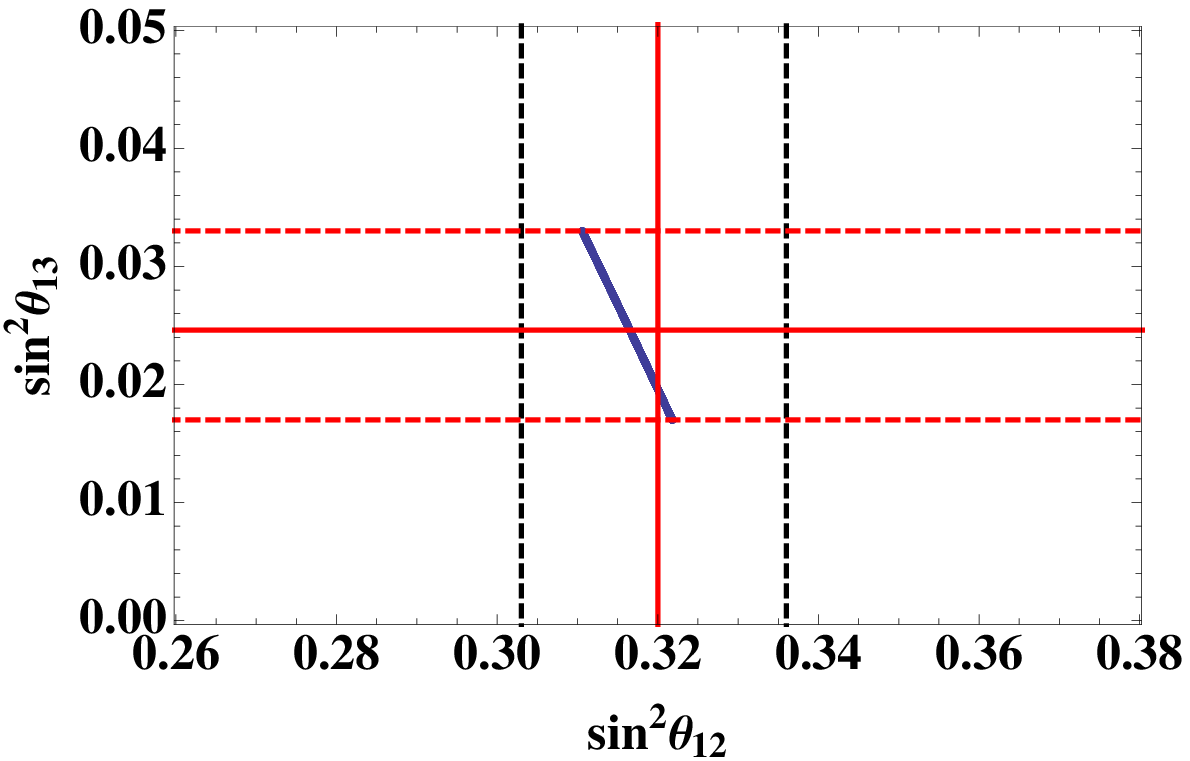}
\caption{The prediction on the $\sin ^2\theta _{12}$--$\sin ^2\theta _{13}$ plane for  NH in Case I. 
The red solid, black dashed, and red dashed lines denote the experimental best fit, 
$1~\sigma $ and $3~\sigma $ level for NH, respectively.}
\label{fig2}
\end{minipage}
\hspace{5mm}
\begin{minipage}[]{0.45\linewidth}
\vskip - 3 mm
\includegraphics[width=7.5cm]{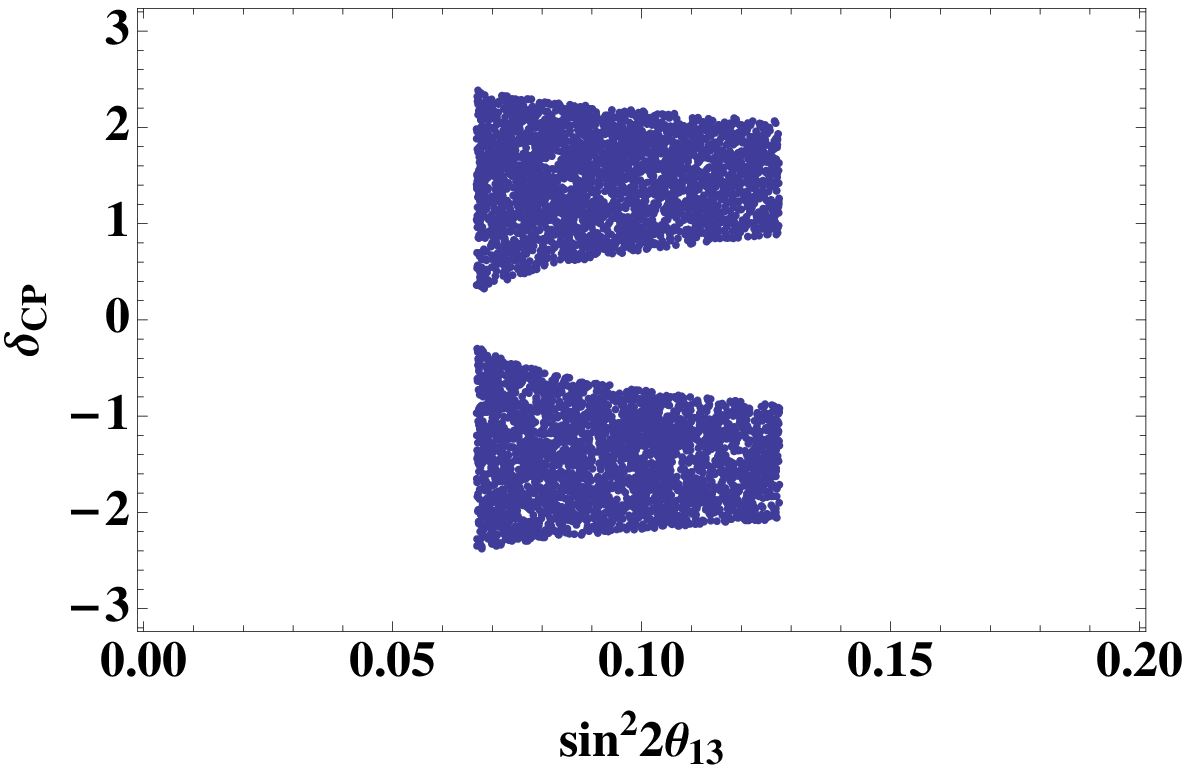}
\caption{The prediction on the $\sin ^22\theta _{13}$--$\delta _{CP}$ plane for NH
in Case I, where $\sin^22\theta_{13}$ is cut by the experimental data.
%The black solid (dashed) line denotes T2K data for NH (IH) with $\sin ^2\theta _{23}=0.5$.  
%The green solid (dashed) ones are for NH (IH) with $\sin^2\theta _{23}=0.4$ (right)
%and  $0.6$ (left).
}
\label{fig3}
\end{minipage}
\end{figure}
%%%%%%%%%%%%%%%%%%%%%%%%%%%%%%%%%%%%%%%%%%%%%
%%%%%%%%%%%%%%%%%%%%%%%%%%%%%%%%%%%%%%%%%%%%%
\begin{figure}[b!]
\begin{minipage}[]{0.45\linewidth}
\includegraphics[width=7.5cm]{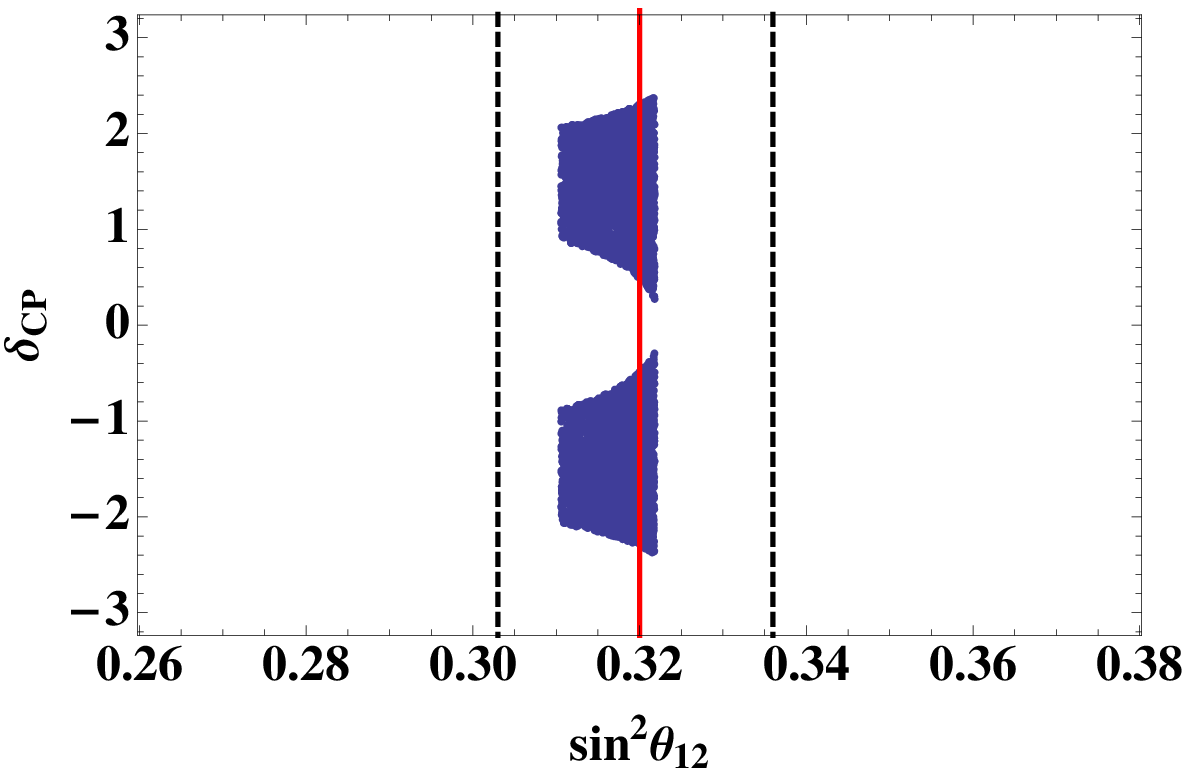}
\caption{The prediction on the $\sin ^2\theta _{12}$--$\delta _{CP}$ plane for  NH 
in Case I. 
The red solid and black dashed lines denote the experimental best fit and
$1~\sigma $ level for NH, respectively.}
\label{fig4}
\end{minipage}
\hspace{5mm}
\begin{minipage}[]{0.45\linewidth}
\includegraphics[width=7.5cm]{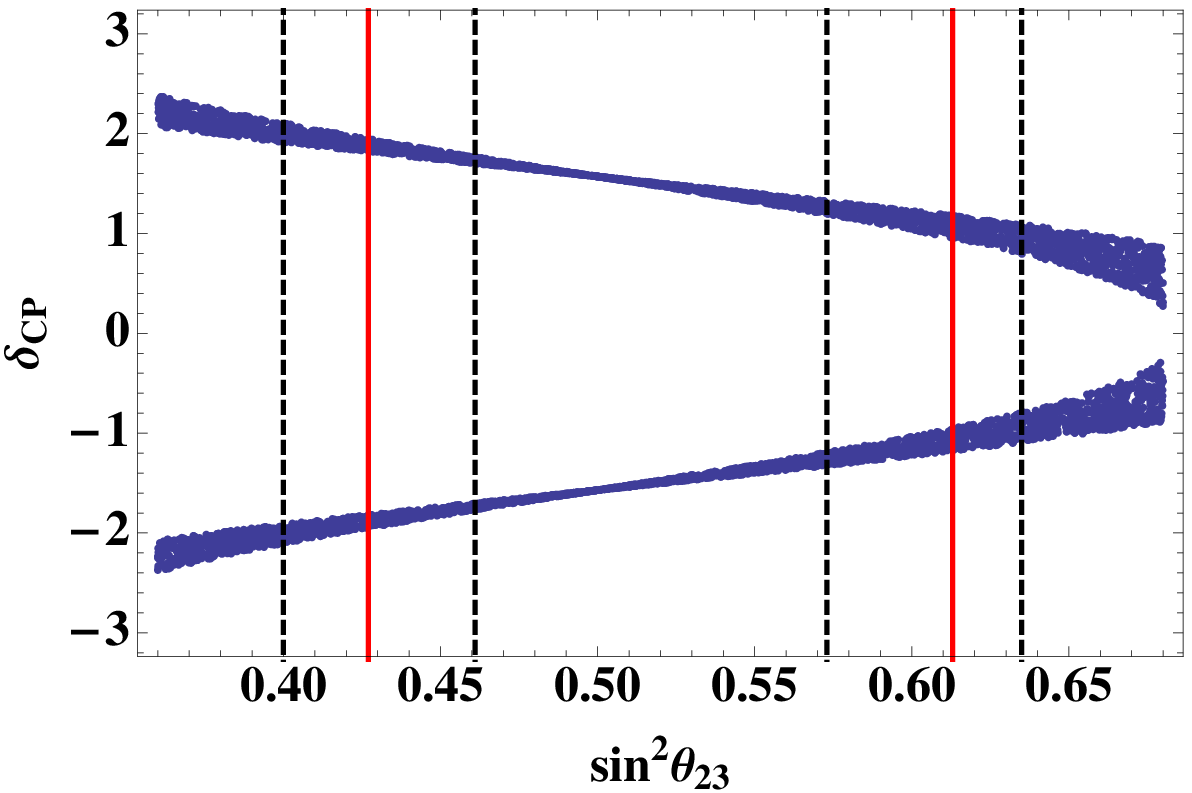}
\caption{The prediction on the $\sin ^2\theta _{23}$--$\delta _{CP}$ plane for  NH 
in Case I. 
The two red solid lines and the black dashed lines denote the experimental best fits and   at 
$1~\sigma $ level, respectively,  for NH.}
\label{fig5}
\end{minipage}
\end{figure}
%%%%%%%%%%%%%%%%%%%%%%%%%%%%%%%%%%%%%%%%%%%%%
%%%%%%%%%%%%%%%%%%%%%%%%%%%%%%%%%%%%%%%%%%%%%
\begin{figure}[t!]
\begin{minipage}[]{0.45\linewidth}
\vskip 0.3 cm
\includegraphics[width=7.5cm]{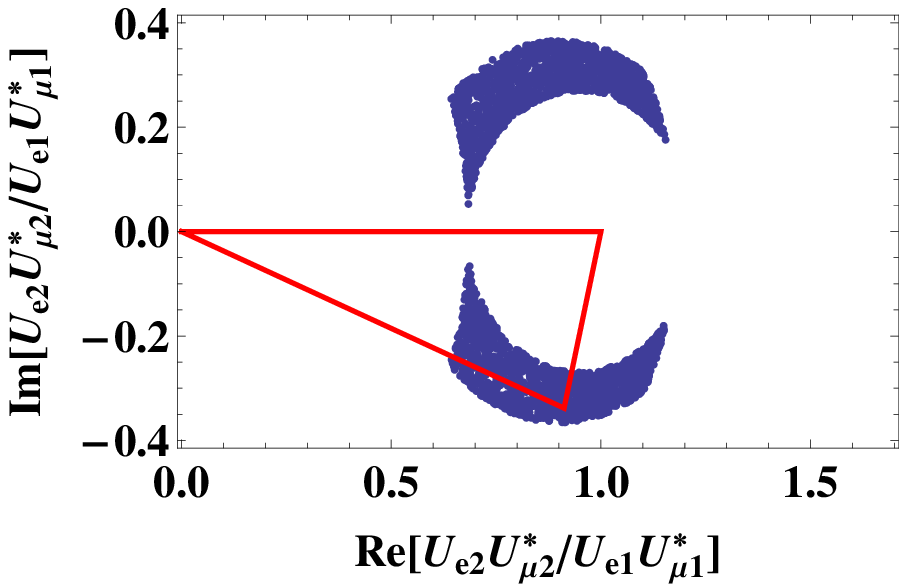}
\caption{The predicted vertex in the  unitarity triangle for  NH  in Case I. 
The red triangle denotes a reference one.}
\label{fig6}
\end{minipage}
\hspace{5mm}
\begin{minipage}[]{0.45\linewidth}
\vskip -0.2 cm
\includegraphics[width=7.5cm]{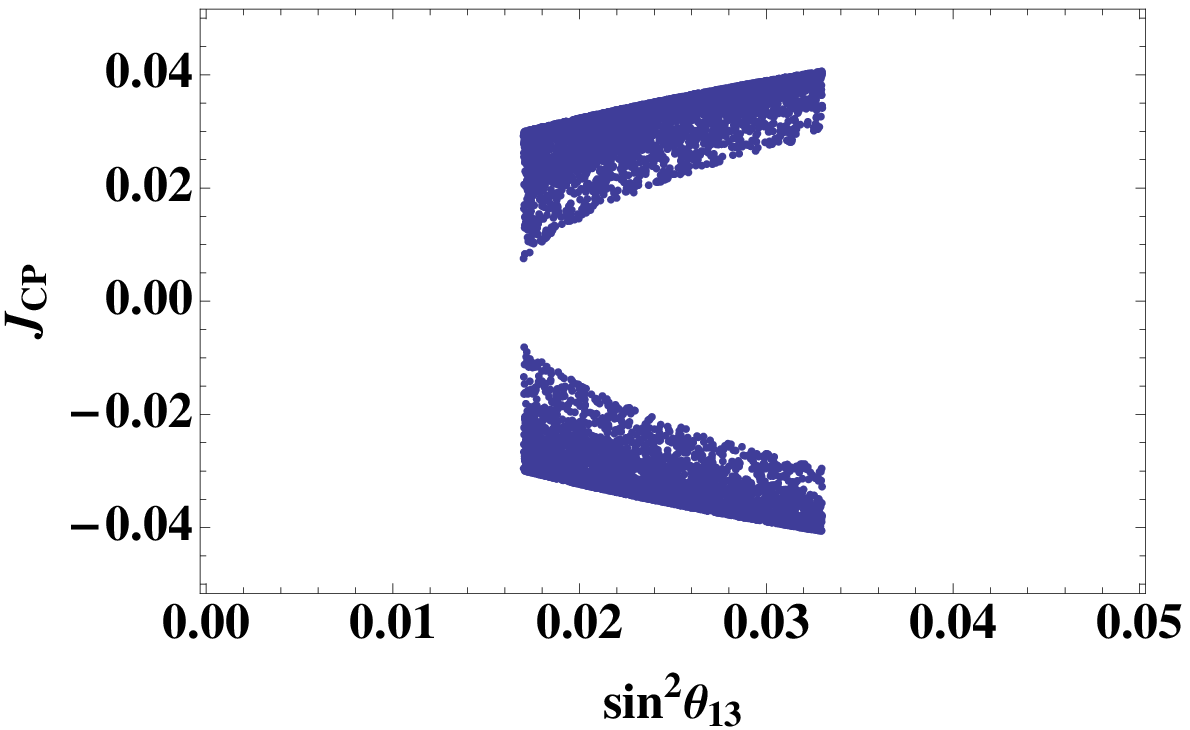}
\caption{The predicted plot of  $J_{CP}$ versus $\sin^2\theta_{13}$ for  NH 
in Case I.}
\label{fig7}
\end{minipage}
\end{figure}
%%%%%%%%%%%%%%%%%%%%%%%%%%%%%%%%%%%%%%%%%%%%%
%%%%%%%%%%%%%%%%%%%%%%%%%%%%%%%%%%%%%%%%%%%%%
%%%%%%%%%%%%%%%%%%%%%%%%%%%%%%%%%%%%%%%%%%%%%
\subsection{Case I}
Let us  discuss  the  NH case, where the data of Eq.(\ref{lepton-mixing-3sigma}) are input. The IH case is only different from the NH one in the input data
 of $\sin ^2\theta _{23}$. 
We start with presenting the result for the Case I, in which  the  $2$-$3$  plane of neutrino generations is rotated  in the TBM basis
as in Eq.(\ref{RO1}).
%%%%%%%%%%%%%%%%%%%%%%%%%%
%%%%%%%%%%%%%%%%%%%%%%%%%%%%%%

We show the prediction on the $\sin ^2\theta _{12}$--$\sin^2\theta _{13}$ plane in Fig.~\ref{fig2}.
It is found that the  predicted linear relation between $\sin ^2\theta _{12}$ and 
$\sin^2\theta _{13}$ is completely consistent with the experimental data.
We obtain $\sin^2\theta _{12}\leq 1/3$ as seen in the sum rule of  Eq.(\ref{sum1}).
Actually, the allowed region of $\sin ^2\theta _{12}$ is restricted by the observed value of 
 $\sin ^2\theta _{13}$ 
such as $0.310<\sin ^2\theta _{12}<0.323$. 
%%%%%%%%%%%%%%%%
%%%%%%%%%%%%%%%%
On the other hand,  since there is no clear  correlation between $\sin^2\theta_{23}$
and $\sin^2\theta _{13}$ as seen in Eq.(\ref{RO1mix}),we omit it in our figures.

Recently, the T2K experiment has presented new data~\cite{Abe:2014ugx}, in which the CP violating Dirac phase is  
% and 
$-1.18\pi <\delta _{CP}<0.15\pi ~(-0.91\pi <\delta _{CP}<-0.08\pi )$ for the NH (IH)
with $\sin ^2\theta _{23}=0.514^{+0.055}_{-0.056}~(0.511\pm 0.055)$.
We show the prediction for the CP violating Dirac phase
versus  $\sin ^22\theta _{13}$ in Fig. \ref{fig3}.
% with the T2K data. 
 %Since the observed  CP violating Dirac phase depends on
% the value of  $\sin ^2\theta _{23}$, we show  the  T2K data for
%each  $\sin^2\theta _{23}$:
% the upper bound for $\sin^2\theta _{23}=0.4$, the central value for $0.5$,
% and lower bound for$0.6$ in Fig.~\ref{fig3}.
The CP violating Dirac phase  is predicted as 
 $0.09\pi\lesssim |\delta _{CP}|\lesssim 0.76\pi$, which
  will be tested  in the future.

In order to see other mixing angle dependences of  $\delta_{CP}$, 
we  show $\delta_{CP}$ versus $\sin ^2\theta _{12}$ and $\sin ^2\theta _{23}$ in Figs.~\ref{fig4} and \ref{fig5}, respectively.  As seen in the sum rules of  Eq.(\ref{sum1}),
our predicted $\delta_{CP}$  depends   on  $\sin ^2\theta _{23}$ clearly,
but does not so on  $\sin ^2\theta _{12}$.
The clear dependence between $\delta_{CP}$ and $\sin ^2\theta _{23}$ is
attributed to the precise measurement of $\sin\theta_{13}$ as seen in Eq.(\ref{sum1}).
The more precise data of $\sin\theta_{13}$ will be  helpful to test this case.

It is found that $\delta_{CP}$ is maximal $\pm\pi/2$ at
 $\sin ^2\theta _{23}=1/2$. 
 If  $\sin ^2\theta _{23}$ is smaller (larger) than $1/2$,
 $|\delta_{CP}|$ is larger (smaller) than  $\pi/2$. 
Thus, we can test our prediction clearly
since the observed value of $\sin ^2\theta _{23}$ is improved in the future experiments.

For the IH case,  the predicted lower bound is reduced to  
$0.15\pi\lesssim |\delta _{CP}|\lesssim 0.73\pi$ since the allowed region of 
$\sin^2\theta_{23}$ is narrow slightly compared with the NH case.
 
Let us discuss the  unitarity triangle of leptons. The unitarity condition 
\begin{equation}
U_{e1}U_{\mu 1}^\ast +U_{e2}U_{\mu 2}^\ast +U_{e3}U_{\mu 3}^\ast =0
\end{equation}
is expressed as a triangle on the complex plane. 
We show  the  predicted vertex of the triangle in Fig.~\ref{fig6}, in which
 the reference  triangle at
\begin{equation}
\delta _{CP}=-\frac{\pi}{2},\quad \sin ^2\theta _{13}=0.0251,\quad 
\sin ^2\theta _{12}=0.312,\quad \sin ^2\theta _{23}=0.514,
\end{equation}
is presented for the eye guide.
Thus, the flavor mixing  will be tested in the unitarity triangle.
 We also show  $J_{CP}$ versus  $\sin ^2\theta _{13}$ in  Fig.~\ref{fig7}.
 The predicted region is $0.005<|J_{CP}|<0.04$. 
 
%%%%%%%%%%%%%%%%%%%%%%%%%%%%%%%%%%%%%%%%%%%%%
\begin{figure}[t!]
%\vskip 1 cm
\begin{minipage}[]{0.45\linewidth}
\vskip 0.5 cm
\includegraphics[width=7.5cm]{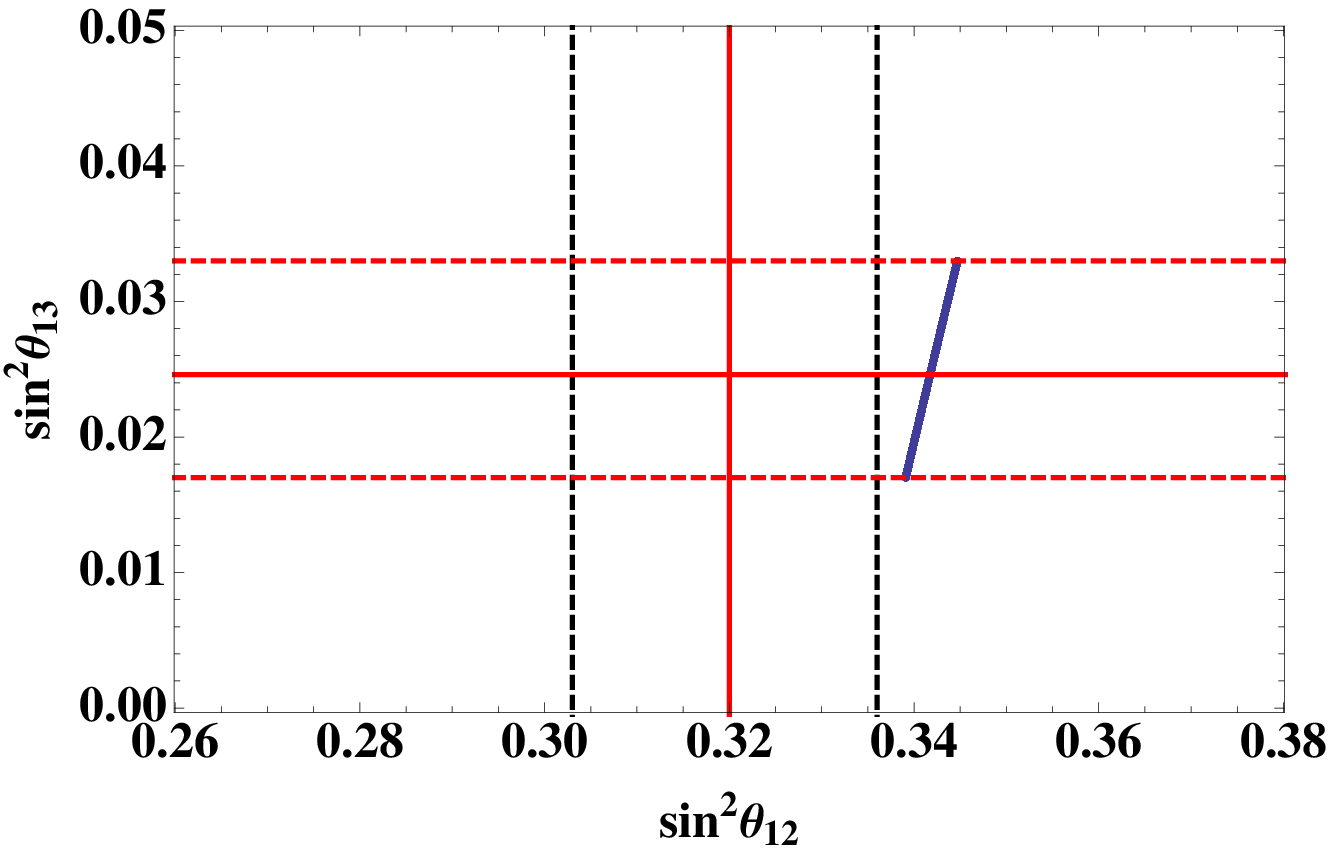}
\caption{The prediction on the $\sin ^2\theta _{12}$--$\sin ^2\theta _{13}$ plane for NH in Case II. 
The red solid, black dashed, and red dashed lines denote the experimental best fit, 
$1~\sigma $ and $3~\sigma $ level for NH, respectively.}
\label{fig8}
\end{minipage}
\hspace{5mm}
\begin{minipage}[]{0.45\linewidth}
\vskip -0.3 cm
\includegraphics[width=7.5cm]{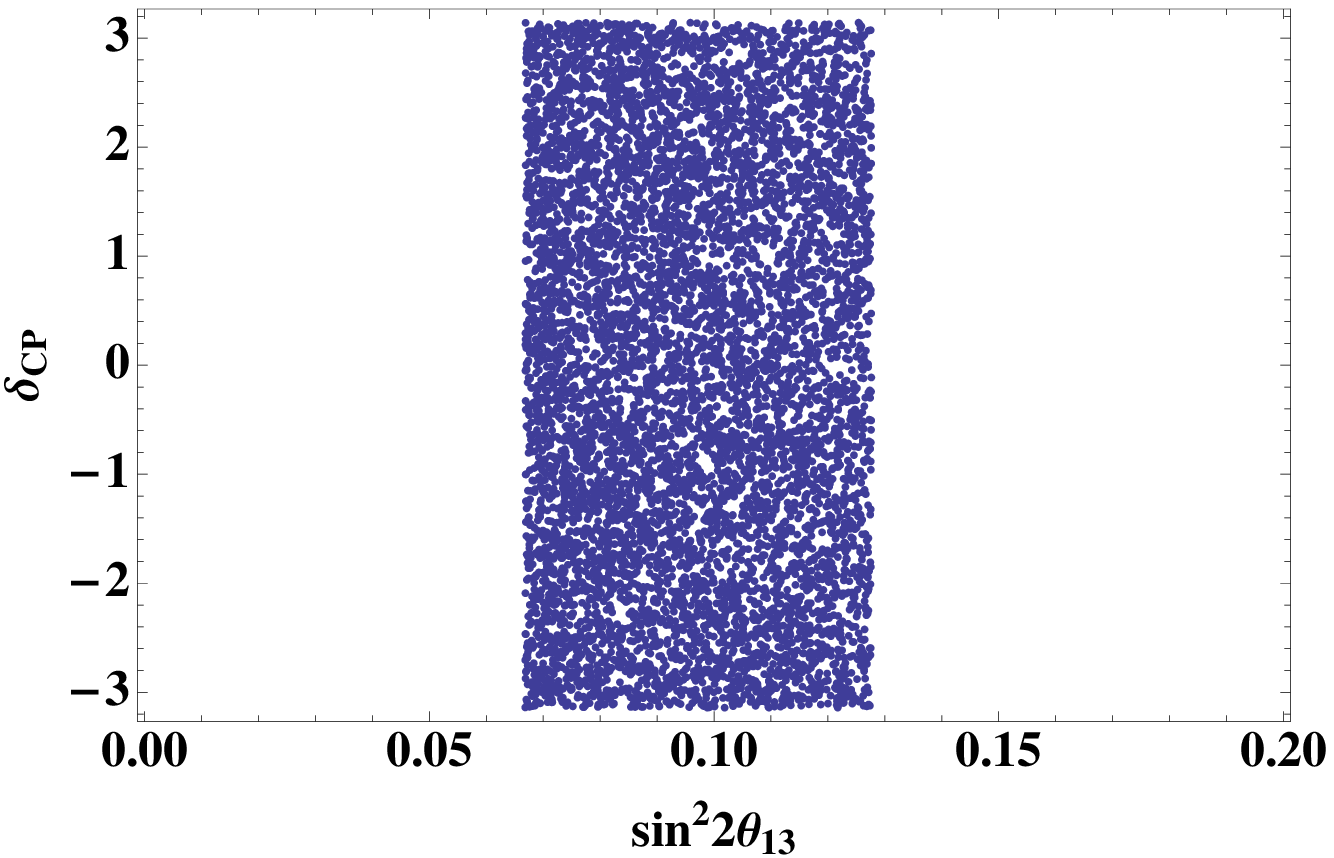}
\caption{The prediction on the $\sin ^22\theta _{13}$--$\delta _{CP}$ plane for  NH
in Case II,  where $\sin^22\theta_{13}$ is cut by the experimental data. 
%The black solid (dashed) line denotes T2K data for NH (IH) with $\sin ^2\theta _{23}=0.5$.  
%The green solid (dashed) ones are for NH (IH) with $\sin^2\theta _{23}=0.4$ (right)
%and  $0.6$ (left).
}
\label{fig9}
\end{minipage}
\end{figure}

%%%%%%%%%%%%%%%%%%%%%%%%%%%%%%%%%%%%%%%%%%%%%

%%%%%%%%%%%%%%%%%%%%%%%%%%%%%%%%%%%%%%%%%%%%%
\begin{figure}[h!]
\begin{minipage}[]{0.45\linewidth}
\includegraphics[width=7.5cm]{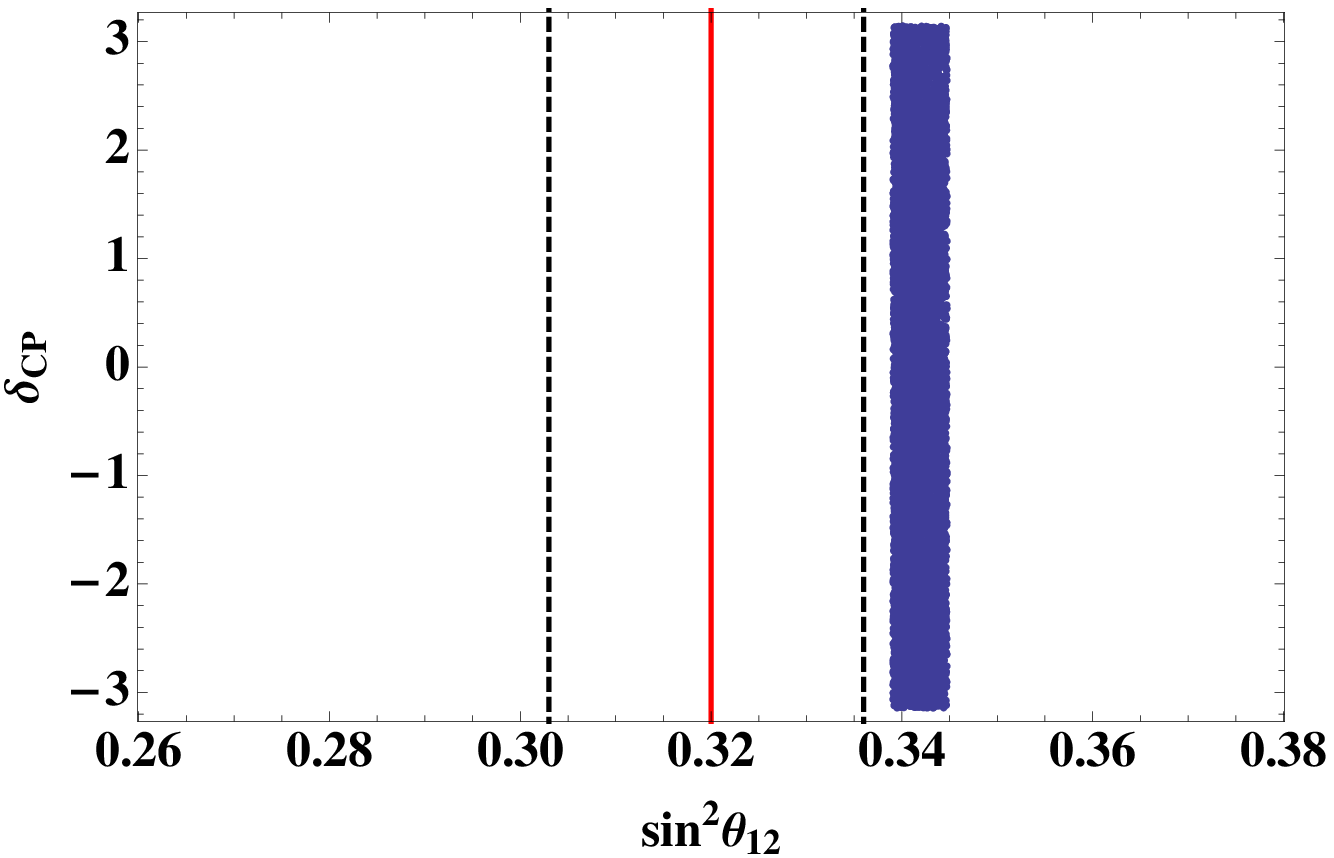}
\caption{The allowed region on the $\sin ^2\theta _{12}$--$\delta _{CP}$ plane for NH in Case II. 
The red solid and black dashed lines denote the experimental best fit and at
$1~\sigma $ level for NH, respectively.}
\label{fig10}
\end{minipage}
\hspace{5mm}
\begin{minipage}[]{0.45\linewidth}
\includegraphics[width=7.5cm]{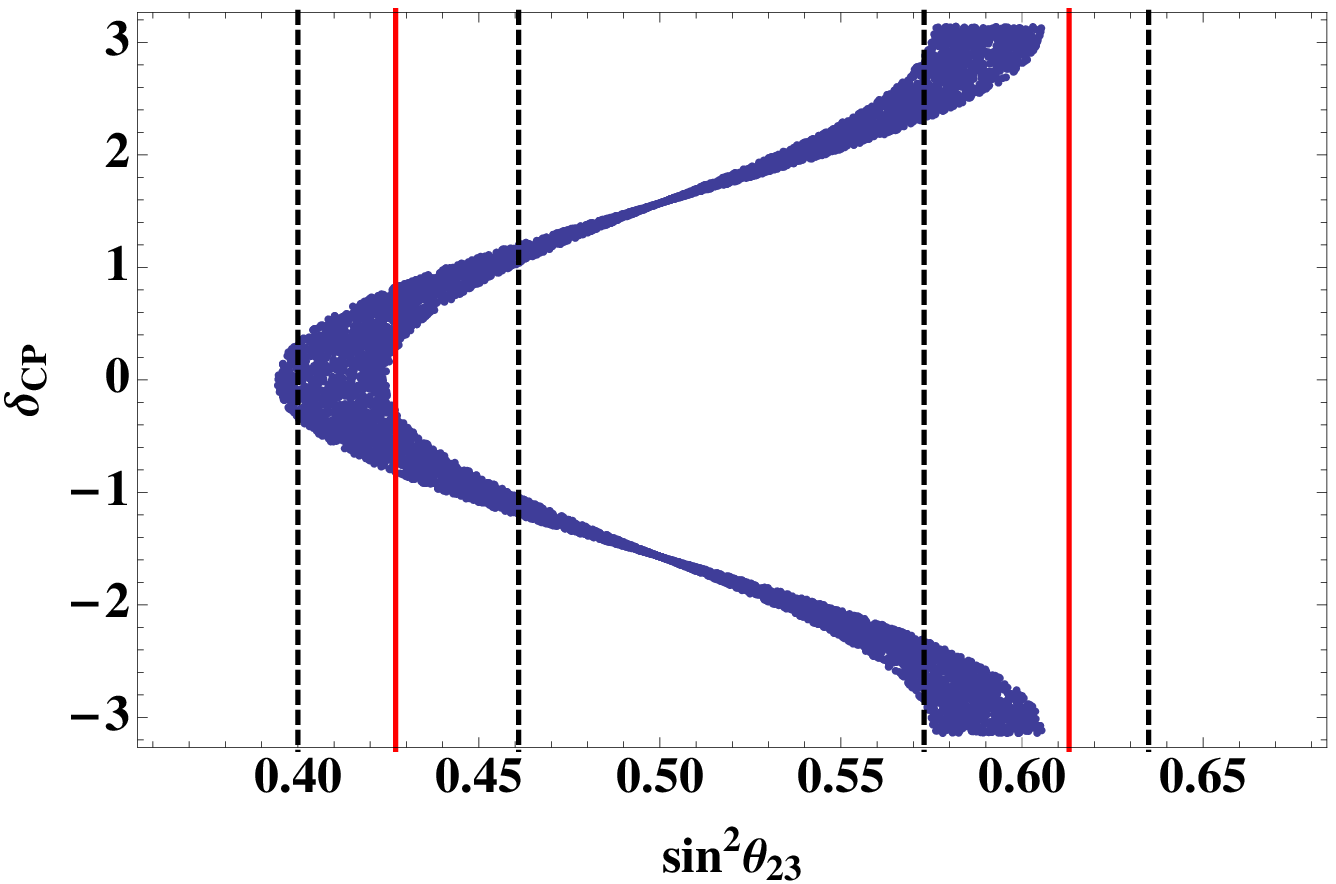}
\caption{The allowed region on the $\sin ^2\theta _{23}$--$\delta _{CP}$ plane for  NH 
in Case II. 
The two red solid lines and the black dashed lines denote the experimental best fits and  at 
$1~\sigma $ level, respectively,  for NH.}
\label{fig11}
\end{minipage}
\end{figure}
%%%%%%%%%%%%%%%%%%%%%%%%%%%%%%%%%%%%%%%%%%%%%
%%%%%%%%%%%%%%%%%%%%%%%%%%%%%%%%%%%%%%%%%%%%%
\begin{figure}[h!]
\begin{minipage}[]{0.45\linewidth}
\includegraphics[width=7.5cm]{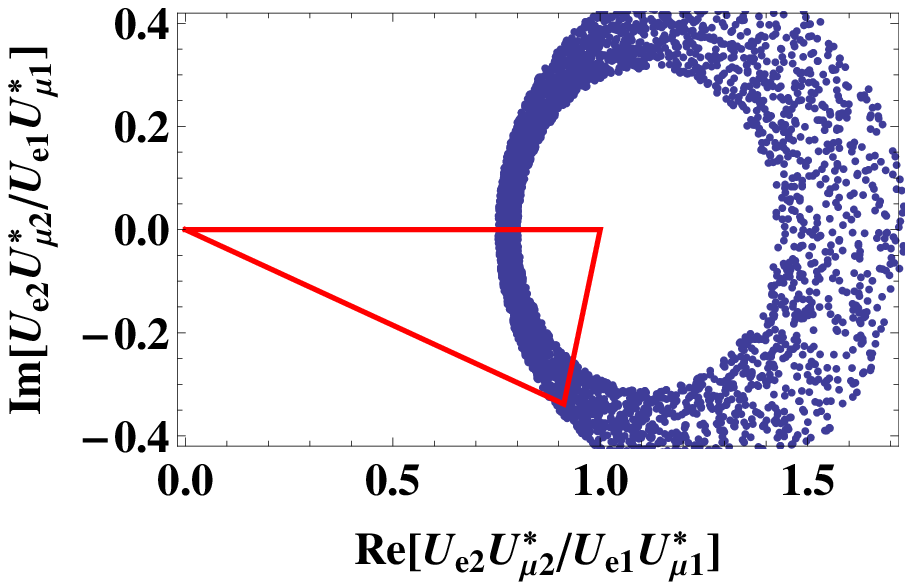}
\caption{The predicted vertex in the  unitarity triangle with  a reference one for  NH in Case II.}
\label{fig12}
\end{minipage}
\hspace{5mm}
\begin{minipage}[]{0.45\linewidth}
\includegraphics[width=7.5cm]{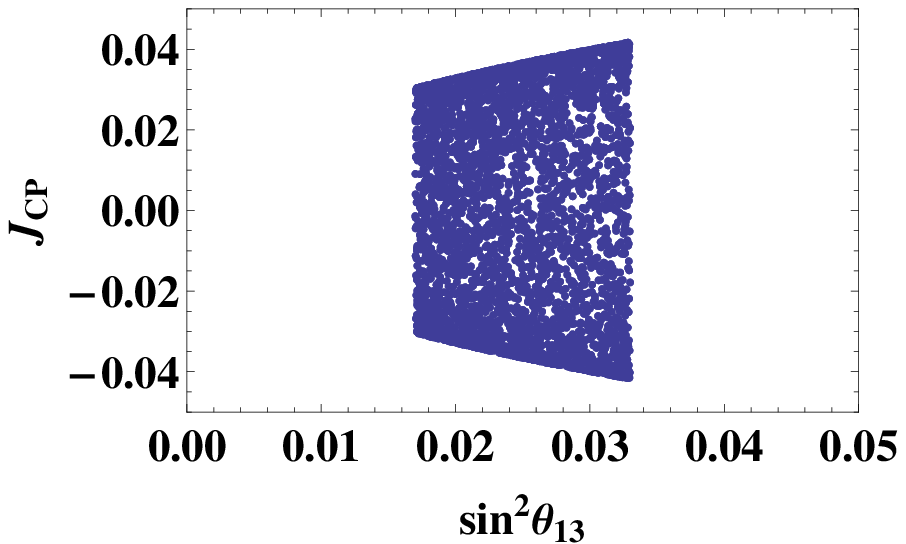}
\caption{The predicted plot of  $J_{CP}$ versus $\sin^2\theta_{13}$ for NH in Case II.}
\label{fig13}
\end{minipage}
\end{figure}
%%%%%%%%%%%%%%%%%%%%%%%%%%%%%%%%%%%%%%%%%%%%%

%%%%%%%%%%%%%%%%%%%%%%%%%%%%%%%%%%%%%%%%%%%%%
\begin{figure}[h!]
\begin{minipage}[]{0.45\linewidth}
\vspace{7mm}
\includegraphics[width=7.5cm]{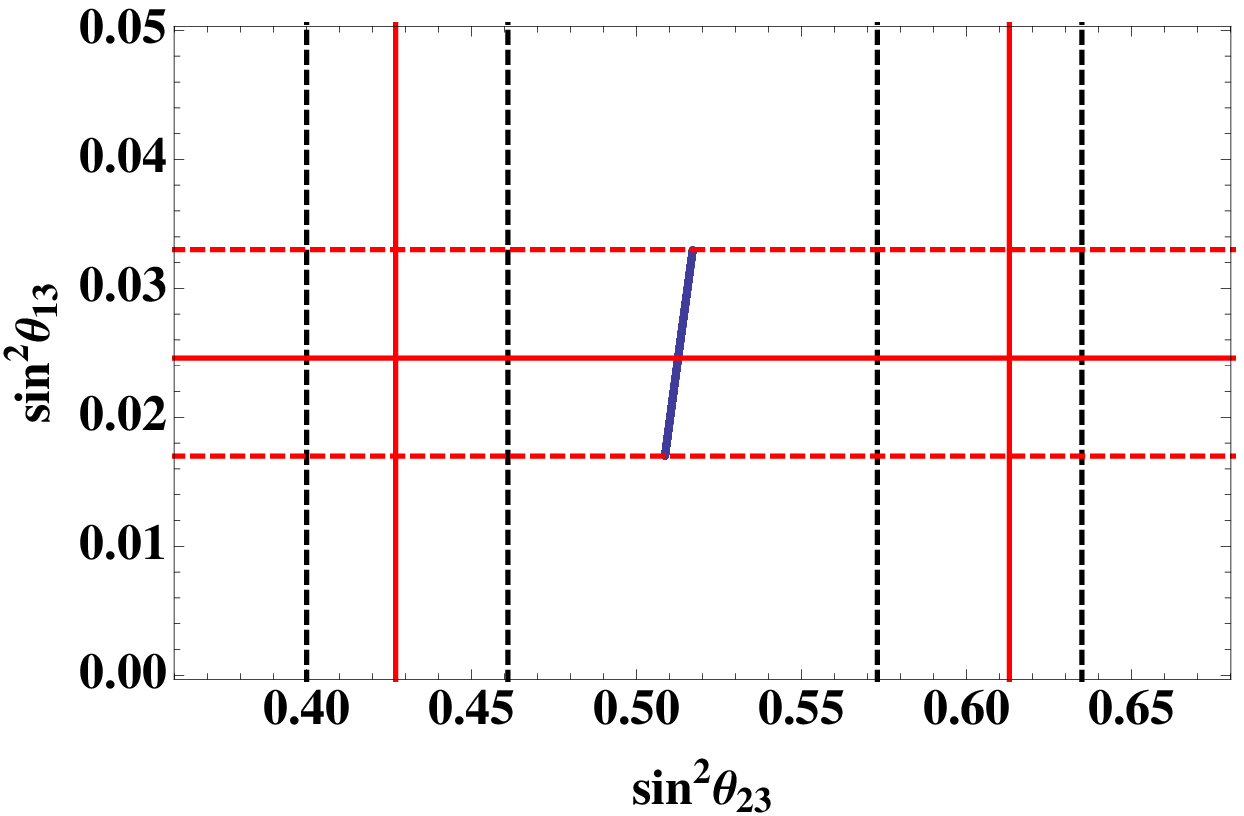}
\caption{The prediction on the $\sin ^2\theta _{12}$--$\sin ^2\theta _{13}$ plane for NH in Case III.
The red solid, black dashed, and red dashed lines denote the experimental best fit, 
$1~\sigma $ and $3~\sigma $ level for NH, respectively.}
\label{fig14}
\end{minipage}
\hspace{5mm}
\begin{minipage}[]{0.45\linewidth}
\vskip -0.1 cm
\includegraphics[width=7.5cm]{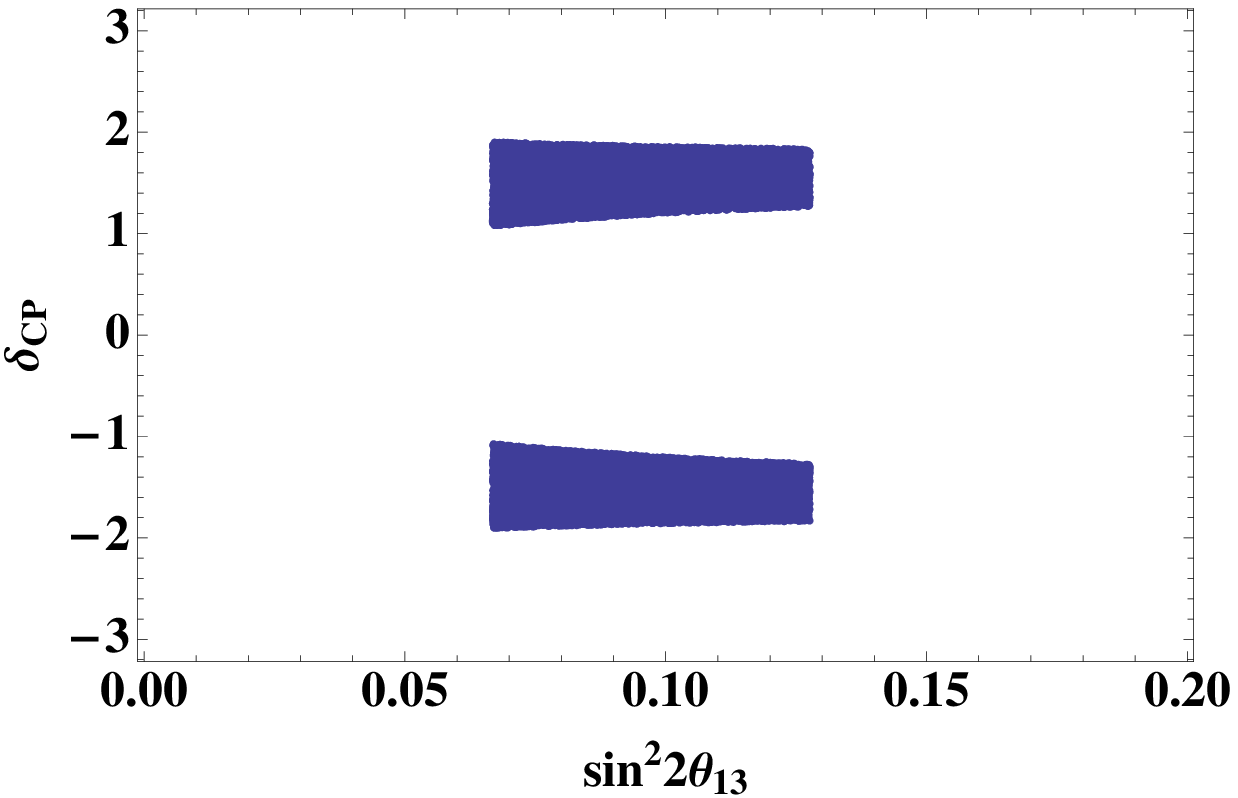}
\caption{The prediction on the $\sin ^22\theta_{13}$--$\delta _{CP}$ plane for NH
in Case III,
 where $\sin^22\theta_{13}$ is cut by the experimental data. 
%The black solid (dashed) line denotes T2K data for NH (IH) with $\sin ^2\theta _{23}=0.5$.  
%The green solid (dashed) ones are for NH (IH) with $\sin^2\theta _{23}=0.4$ (right)
%and  $0.6$ (left).
}
\label{fig15}
\end{minipage}
\end{figure}
%%%%%%%%%%%%%%%%%%%%%%%%%%%%%%%%%%%%%%%%%%%%%

\subsection{Case II}
We discuss the Case II, 
in which the $1$-$3$  plane of neutrino generations is rotated  in the TBM basis as 
in Eq.(\ref{RO2}). 
We show the predicted line on the $\sin^2\theta_{12}$--$\sin^2\theta_{13}$ plane in  Fig.~\ref{fig8}. 
In this case, the predicted $\sin ^2\theta _{12}$ is outside of $1~\sigma $ of the experimental data, which is contrast with the one in the   Case I. 
We obtain   $\sin ^2\theta _{12}\geq 1/3$ as seen in the sum rule of Eq.(\ref{sum2}).

We show the predicted $\delta _{CP}$ versus  $\sin ^22\theta _{13}$
 in Fig.~\ref{fig9}. 
 The predicted $\delta _{CP}$ is allowed in all region of $-\pi\sim \pi$.
 In order to see other mixing angle dependences of $\delta _{CP}$,
we show the predicted  $\delta _{CP}$ versus
$\sin ^2\theta _{12}$ and   $\sin ^2\theta _{23}$ 
in  Figs.~\ref{fig10} and \ref{fig11}, respectively.
There is no $\sin^2\theta _{12}$ dependence for the predicted  $\delta _{CP}$
as seen in  Eq.(\ref{RO2mix}).
On the other hand, there is a remarkable $\sin ^2\theta _{23}$ dependence
for $\delta _{CP}$  due to the precise data of $\sin\theta_{13}$ 
as seen in the sum rule of   Eq.(\ref{sum2}).
It is noticed that $\delta_{CP}$ is maximal, $\pm\pi /2$ at
 $\sin ^2\theta _{23}=1/2$ as well as in Case I. 
If $\sin ^2\theta _{23}$ is smaller (larger) than $1/2$,
 $|\delta_{CP}|$ is smaller (larger) than  $\pi/2$. 
Thus, we can test our prediction in the future
since the observed value of $\sin ^2\theta _{23}$ is improved in the future experiments
as well as in Case I.
It is also noted that the $\sin ^2\theta _{23}$ is  predicted as $0.39\lesssim \sin ^2\theta _{23}\lesssim 0.61$,
which is  within the  experimental bound of $3\sigma $ level,
$0.36<\sin ^2\theta _{23}<0.68$.

In Fig.~\ref{fig12}, we show the vertex in the unitarity triangle, which is still allowed in the wide region.
We also show  the Jarlskog invariant $J_{CP}$ versus $\sin^2\theta_{13}$
in  Fig.~\ref{fig13}.  The $J_{CP}$  is predicted to be in the region of $-0.04\lesssim J_{CP}\lesssim 0.04$ . 

For the IH case,  the predictions are the same as the ones in the NH case,
because the predicted region of $\sin^2\theta_{23}$ is inside of the experimental data
of $3\sigma$ as seen in Fig.11.

%%%%%%%%%%%%%%%%%%%%%%%%%%%%%%%%%%%%%%%%%%%%%

%%%%%%%%%%%%%%%%%%%%% Charged Lepton %%%%%%%%%%%%%%

\subsection{Case III}
We discuss Case III, in which the 1-3 plane of   charged lepton
generations is rotated as in Eq.(\ref{RO3}). 
We show the prediction on the $\sin^2\theta_{23}$--$\sin^2\theta_{13}$
 plane in  Fig.~\ref{fig14}. 
 The predicted $\sin ^2\theta _{23}$ is outside of $1~\sigma $ of the experimental data.
The predicted   $\sin ^2\theta _{23}$ is slightly larger than  $1/2$ as seen   in the sum rule of Eq.(\ref{sum3}).
The predicted  $\sin ^2\theta _{13}$ is sensitive to the magnitude of 
 $\sin ^2\theta _{23}$. 
On the other hand,  there is no correlation
  between $\sin ^2\theta _{12}$ and $\sin ^2\theta _{13}$.
This is contrast to Case I and Case II.
The figure is omitted in this paper.

\begin{figure}[t!]
\begin{minipage}[]{0.45\linewidth}
\includegraphics[width=7.5cm]{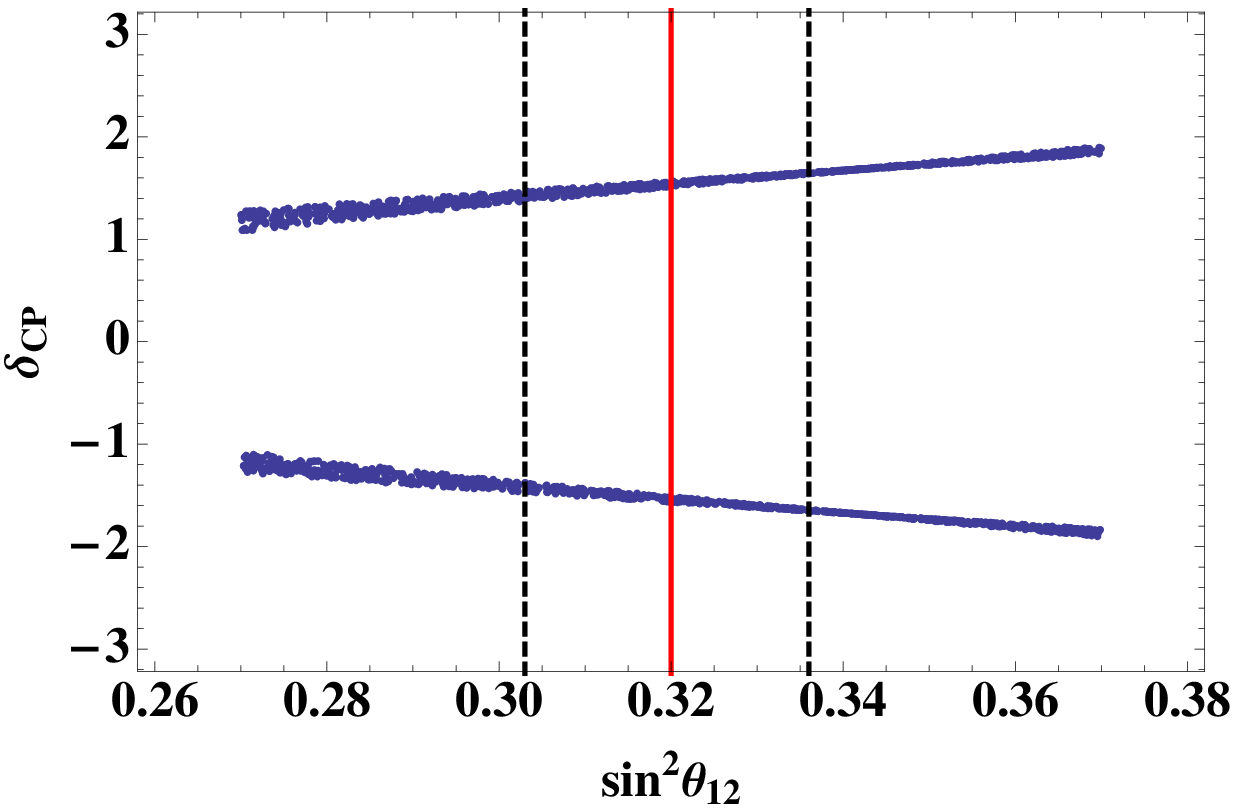}
\caption{The allowed region on the $\sin ^2\theta _{12}$--$\delta _{CP}$ plane for  NH 
in Case III. 
The red solid and black dashed lines denote the experimental best fit and at
$1~\sigma $ level for NH, respectively.}
\label{fig16}
\end{minipage}
\hspace{5mm}
\begin{minipage}[]{0.45\linewidth}
\includegraphics[width=7.5cm]{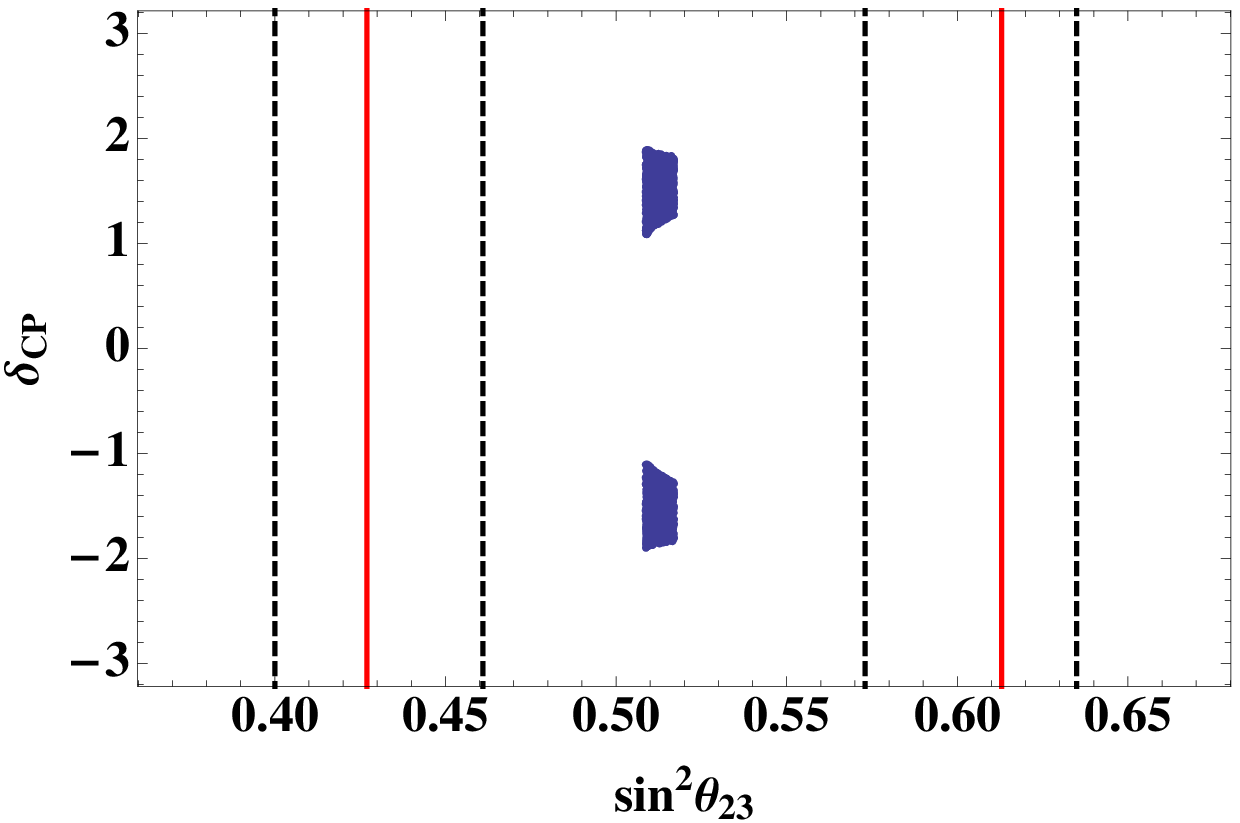}
\caption{The allowed region on the $\sin ^2\theta _{23}$--$\delta _{CP}$ plane for  NH 
in Case III. 
The two red solid lines and the black dashed lines denote the experimental best fits and   at 
$1~\sigma $ level, respectively,  for NH.}
\label{fig17}
\end{minipage}
\end{figure}

We show the scatter plot of $\delta _{CP}$  versus  $\sin ^22\theta _{13}$
in  Fig.~\ref{fig15}.
The $\delta _{CP}$ is predicted to be 
$\delta _{CP}\simeq \pm (0.35\pi\sim 0.60\pi)$, 
which  is testable in the  future experiments.
We show $\delta _{CP}$ versus   $\sin^2\theta _{12}$ and $\sin^2\theta _{23}$
in Figs.~\ref{fig16} and \ref{fig17}, respectively. 
There is a clear $\sin^2\theta_{12}$ dependence for  $\delta _{CP}$
due to the precise data of $\sin\theta_{13}$, 
which is understood in the sum rule of Eq.(\ref{sum3}).
It is remarked that $\delta_{CP}$ is maximal value $\pm\pi/2$ 
at  $\sin^2\theta_{12}\simeq 1/3$. 
If  $\sin ^2\theta _{12}$ is smaller (larger) than $1/3$,
 $|\delta_{CP}|$ is smaller (larger) than  $\pi/2$. 
Thus, we can test our prediction 
by the precise data of  $\sin ^2\theta _{12}$  in the future experiments.

\begin{figure}[t!]
\begin{minipage}[]{0.45\linewidth}
\includegraphics[width=7.5cm]{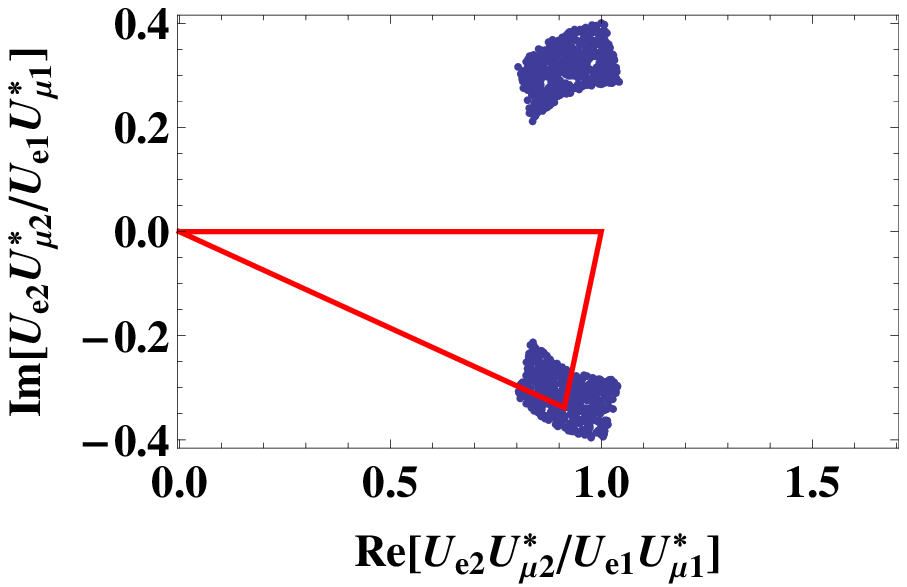}
\caption{The predicted vertex in the unitarity triangle for  NH 
in Case III.}
\label{fig18}
\end{minipage}
\hspace{5mm}
\begin{minipage}[]{0.45\linewidth}
\includegraphics[width=7.5cm]{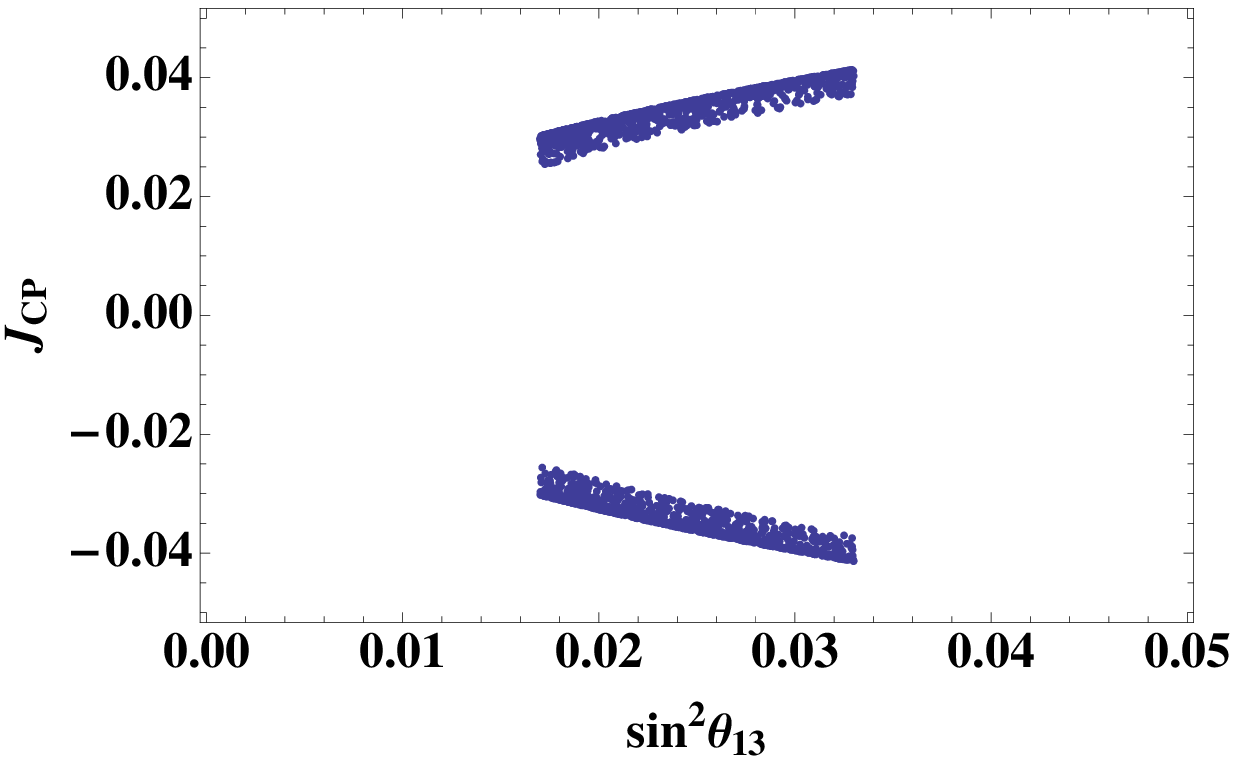}
\caption{The predicted plot of  $J_{CP}$ versus $\sin^2\theta_{13}$ for  NH 
in Case III.}
\label{fig19}
\end{minipage}
\end{figure}

In Fig.~\ref{fig18}, we show the unitarity triangle. The predicted region of the vertex
 is narrow compared with  the ones in  Case I and  Case II. 
 We also show  the Jarlskog invariant $J_{CP}$ versus $\sin^2\theta_{13}$
in Fig.~\ref{fig19}.  The $J_{CP}$  is predicted to be in the narrow region of
 $J_{CP}$ as $0.025\lesssim |J_{CP}|\lesssim 0.04$.
 
 For the IH case,  the predictions are the same as the one  in the NH case
 as well as the Case II,
because the predicted region of $\sin^2\theta_{23}$ is near the maximal one
as seen in Fig. 16. 

%%%%%%%%%%%%%%%%%%%%%%%%%%%%%%%%%%%%%%%%%%%
%%%%%%%%%%%%%%%%%%%%%%%%%%%%%%%%%%%%%%%%%%%%%
%%%%%%%%%%%%%%%%%%%%%%%%%%%%%%%%%%%%%%%%%%%%%
%%%%%%%%%%%%%%%%%%%%%%%%%%%%%%%%%%%%%%%%%%%%%

\subsection{Case IV}
For  Case IV, in which  the 1-2 plane of the  charged lepton generations 
is rotated as in Eq.(\ref{RO4}), 
  we obtain the similar results in the Case III.
 The predictions are obtained by
 replacing $\sin^2\theta _{23}$ with $\cos^2\theta _{23}$ in Case III.
 This situation is easily understood by
 comparing with  Eqs.(\ref{RO3mix2}) and (\ref{RO4mix2}),
 or the sum rules  Eqs.(\ref{sum3}) and (\ref{sum4}).
  Therefore, we omit figures of numerical results.
  
 Instead, we add some comments as follows. 
  The predicted $\sin ^2\theta _{23}$ is slightly  smaller than $1/2$ as seen in 
the sum rule of  Eq.(\ref{sum4}).
 The CP violating Dirac phase  $\delta _{CP}$ depends on
  $\sin^2\theta_{12}$ as well as in Case III, however 
 if  $\sin ^2\theta _{12}$ is smaller (larger) than $1/3$,
 $|\delta_{CP}|$ is larger  (smaller) than  $\pi/2$,
 which is contrast to the one in Fig. \ref{fig16}.
Thus, the precise data of  $\sin ^2\theta _{12}$  can distinguish
  Case III and Case IV in the future.

Finally, we discuss the specific case of Case IV, in which  $\theta_{13}$ is related with the Cabibbo angle
such as  $\sin\theta_{13}=\lambda/\sqrt{2}\simeq 0.16$. This relation  can be derived from 
 the framework of GUT, where the Yukawa matrices for the charged leptons and  the down-type quarks have the same origin.
 We obtain the magnitude of the  CP violating phase  to be 
in the  narrow region $\pm (0.41\pi\sim 0.62\pi)$,
which may prefer the maximal CP violation, $\delta_{CP}=\pm \pi/2$.

\section{Summary}
 We examine  non-trivial relations among  the mixing angles and the CP violating Dirac phase in the typical  four cases of the deviation from the TBM.
The first two cases are given by the  additional rotation of the 2-3 or 1-3 generations of neutrinos in the TBM basis.
Other two cases are given 
by the additional rotation of the 1-3 or 1-2 generations of  charged leptons with  the TBM neutrinos.
These four cases give  different predictions  among three mixing angles and the CP violating Dirac phase.

The rotation of the 2-3 generations of neutrinos in the TBM basis
 predicts $\sin ^2\theta _{12}<1/3$ and  the CP violating Dirac phase  $\delta _{CP}$ to be   $\pm (0.09\pi\sim 0.76\pi)$ for the NH case.
  It has a clear  $\sin ^2\theta _{23}$ dependence, which will be testable in the future.
If $\sin ^2\theta _{23}$ is smaller (larger) than $1/2$,
 $|\delta_{CP}|$ is larger (smaller) than  $\pi/2$. 
 For the IH case,  the predicted  bounds are  reduced to 
$0.15\pi\lesssim |\delta _{CP}|\lesssim 0.73\pi$.
 
The  rotation of the 1-3 generations of neutrinos in the TBM basis
 gives $\sin ^2\theta _{12}>1/3$.  The CP violating Dirac phase
 $\delta _{CP}$ is not constrained by the input of  the present experimental data.
  It has also a remarkable  $\sin ^2\theta _{23}$ dependence.
If $\sin ^2\theta _{23}$ is smaller (larger) than $1/2$,
 $|\delta_{CP}|$ is smaller (larger) than  $\pi/2$. 
 
In the case of the  rotation of the 1-3 and 2-3  charged lepton generations,
 $\sin ^2\theta _{23}$ is slightly  larger  than $1/2$ (Case III) or smaller  than $1/2$ (Case IV).
The CP violating Dirac phase is predicted in the narrow range of 
$\pm(0.35\pi\sim 0.60\pi)$ for both cases,
which may be preferred in the recent T2K experiment.
There is a clear $\sin^2\theta_{12}$ dependence for 
 $\delta _{CP}$, which is contrast to Case I and Case II.
In Case III, 
 $|\delta_{CP}|$ is smaller (larger) than  $\pi/2$
if  $\sin ^2\theta _{12}$ is smaller (larger) than $1/3$. 
In Case IV, 
 $|\delta_{CP}|$ is smaller (larger) than  $\pi/2$
if  $\sin ^2\theta _{12}$ is larger (smaller) than $1/3$. 
Numerical results are same for the NH and IH cases.

 Finally,  in the special case of Case IV,   $\theta_{13}$ is related with the Cabibbo angle
such as  $\sin\theta_{13}=\lambda/\sqrt{2}\simeq 0.16$. This relation  can be derived from 
 the framework of GUT, where the Yukawa matrices for the charged leptons and for the down-type quarks have the same origin.
 The CP violating phase is predicted  to be in the region of 
$\pm (0.41\pi\sim 0.62\pi)$, which prefer the maximal CP violation.

Thus, the CP violating Dirac phase can distinguish the lepton flavor mixing patterns  
at T2K and NO$\nu$A experiments in the future.

%%%%% acknowledgment %%%%%
\vspace{0.4 cm}
\noindent
{\bf Acknowledgment}

We thank W.~Rodejohann for useful comments. 
Y.S. is supported by JSPS Postdoctoral Fellowships for Research Abroad, No.20130600.
M.T. and K.Y.  are  supported by JSPS
 Grand-in-Aid for Scientific Research,  No.24654062 and No.25-5222, respectively.

%%%%%%%%%%%%%%%%%%%%%%%%%%%%%%%%%%%%

\end{document}